\documentclass[journal,10pt]{IEEEtran}
\usepackage{amssymb}
\usepackage{amsmath}
\usepackage{cite}
\usepackage{url}
\usepackage{xcolor}
\usepackage{cite,graphicx,amsmath,amssymb}
\usepackage{subfigure}
\usepackage{citesort}
\usepackage{fancyhdr}
\usepackage{mdwmath}
\usepackage{mdwtab}
\usepackage{caption}
\usepackage{amsthm}
\usepackage{setspace}
\usepackage{algorithm}
\usepackage{algorithmic}
\usepackage{makecell}
\usepackage{diagbox}
\usepackage{stfloats}
\usepackage{balance} 
\usepackage{graphicx}
\usepackage{epstopdf}
\usepackage{epsfig}

\newtheorem{theorem}{Theorem}

\newtheorem{lemma}{Lemma}

\newtheorem{corollary}{Corollary}

\allowdisplaybreaks
\makeatletter
\def\ScaleIfNeeded{
\ifdim\Gin@nat@width>\linewidth \linewidth \else \Gin@nat@width
\fi } \makeatother
 
\begin{document}
\title{Movable-Element STARS-Assisted Near-Field Wideband Communications}
\author{
	Guangyu~Zhu,
	Xidong~Mu,
	Li~Guo,
	Ao~Huang,
	Shibiao~Xu
	\thanks{Part of this work has been submitted to the IIEEE/CIC International Conference on Communications in China, Shanghai, China, August 10–13 2025 \cite{Zhu_conference}.}
	\thanks{Guangyu Zhu, Li Guo, Ao Huang and Shibiao Xu are with the Key Laboratory of Universal Wireless Communications, Ministry of Education, Beijing University of Posts and Telecommunications, Beijing 100876, China, also with the School of Artificial Intelligence, Beijing University of Posts and Telecommunications, Beijing 100876, China, and also with the National Engineering Research Center for Mobile Internet Security Technology, Beijing University of Posts and Telecommunications, Beijing 100876, China (email:\{Zhugy, guoli, huangao, shibiaoxu\}@bupt.edu.cn).}
	\thanks{Xidong Mu is with the Centre for Wireless Innovation (CWI), Queen's University Belfast, Belfast, BT3 9DT, U.K. (e-mail:
		x.mu@qub.ac.uk).}
}
\maketitle
\begin{abstract}
	A novel movable-element simultaneously transmitting and reflecting surface (ME-STARS)-assisted near-field wideband communication framework is proposed. In particular, the position of each STARS element can be adjusted to combat the significant wideband beam squint issue in the near field instead of using costly true-time delay components. Four practical ME-STARS element movement modes are proposed, namely region-based (RB), horizontal-based (HB), vertical-based (VB), and diagonal-based (DB) modes. Based on this, a near-field wideband multi-user downlink communication scenario is considered, where a sum rate maximization problem is formulated by jointly optimizing the base station (BS) precoding, ME-STARS beamforming, and element positions. To solve this intractable problem, a two-layer algorithm is developed. For the inner layer, the block coordinate descent optimization framework is utilized to solve the BS precoding and ME-STARS beamforming in an iterative manner. For the outer layer, the particle swarm optimization-based heuristic search method is employed to determine the desired element positions. Numerical results show that: 1) the ME-STARSs can effectively address the beam squint for near-field wideband communications compared to conventional STARSs with fixed element positions; 2) the RB mode achieves the most efficient beam squint effect mitigation, while the DB mode achieves the best trade-off between performance gain and hardware overhead; and 3) an increase in the number of ME-STARS elements or BS subcarriers substantially improves the system performance.
\end{abstract}
\begin{IEEEkeywords}
	Simultaneously transmitting and reflecting surfaces, near-field communication, wideband beam squint, movable element.
\end{IEEEkeywords}
\section{Introduction}
High-frequency communications, especially in the millimeter-wave (mmWave) and terahertz (THz) bands, are expected to play a pivotal role in the development of the sixth-generation (6G) wireless networks \cite{6G_high-frequency}. As demand for ultra-high data rates, low latency, and ubiquitous connectivity grows, ultra-high-frequency bands offer the wide bandwidths necessary to support advanced 6G applications, including immersive augmented reality, holographic communication, and large-scale Internet-of-Things (IoT) networks \cite{6G_application}. However, mmWave and THz communications face significant propagation challenges, including high path loss and sensitivity to block, which degrade the efficiency and reliability of data transmission. As a result, enhancing signal quality and expanding coverage in high-frequency communications have become critical research areas.

With its ability to smartly \emph{adjust} wireless transmission environments, reconfigurable intelligent surfaces (RISs) have emerged as a promising technology for future wireless communication networks \cite{RIS_survey}. In particular, RISs are viewed as a key enabler for enhancing both spectral and energy efficiency in mmWave and THz communication scenarios \cite{Huang_EE}. However, due to their reflecting-only characteristic, reflective RISs can only serve users within the \emph{half space}, which limits their deployment benefits. To address this limitation, a novel technology known as simultaneously transmitting and reflecting surfaces (STARSs), which incorporates both reflective and transmissive capabilities, has been proposed \cite{Mu_star}. Compared to conventional RISs, STARSs extend coverage to \emph{full space} and offer three flexible operating protocols: energy splitting (ES), mode selection (MS), and time switching (TS) \cite{Mu_survey}. This versatility enables STARSs to unlock new possibilities and tackle emerging challenges in mmWave and THz communications.

To overcome severe path loss in mmWave and THz communications, large STARS aperture sizes are essential. However, as STARS aperture sizes increase and communication frequencies continue to rise, the Rayleigh distance in STARS-assisted systems expands from just a few meters to tens or even hundreds of meters. In this scenario, the near-field effect becomes increasingly significant and can no longer be ignored \cite{Liu_NF_survey}. Unlike far-field communications, where signals are steered in a direction like a flashlight \cite{Far_field_beamforming}, near-field communications can enable enhanced spatial resolution and precise energy focusing to a specific location like a spotlight, thus offering significant advantages for communication system design \cite{Near_field_beamforming,Zhang_NF}. Consequently, this unique transmission characteristic also introduces emerging distance domain considerations, alongside the traditional angle domain, into the STARS beamforming design, adding complexity and challenge to the design process \cite{Xu_STAR-RIS_NF,Mu_RIS_NF}.
Especially in THz communications with abundant bandwidth resources, these challenges are further compounded due to the wideband \emph{beam squint} effect \cite{wideband_BS}. This is because the channel array response in wideband systems varies with frequency, whereas the STARS is inherently restricted to only performing frequency-independent beamforming. As a result, beams on different subcarriers are dispersed and refocused at different locations after reflection or transmission, leading to a reduction in transmission efficiency \cite{Zhang_NF_wideband}. Although the beam squint issue also occurs in far-field wideband communications, it is more pronounced in near-field wideband scenarios, where the beam may become misfocused relative to the user's position in both the distance and angle domains. Accordingly, addressing these issues in near-field wideband communications has recently become a critical research focus.
\subsection{Prior Works}
\emph{1) Conventional Near-field Wideband Communications:} In recent years, significant research efforts have been dedicated to addressing the near-field beam squint effect caused by phased arrays at base stations (BS) or access points. For instance, the authors in \cite{Wang_wideband} introduced a sub-connected true-time delay (TTD) architecture at the BS equipped with a large-scale uniform linear antenna (ULA) array, demonstrating its effectiveness in suppressing the beam squint effect in near-field wideband communications. Their results validated that this architecture achieves a favorable tradeoff between performance gains and hardware complexity. Building on this, the authors of \cite{Guo_NF_wideband} extended the TTD-based architecture to a circular antenna array at the BS and proposed two wideband beamforming optimization schemes to effectively mitigate the beam squint effect in near-field wideband communications. Additionally, a phase-delay focusing method employing delay phase precoding architecture was proposed to overcome the beam squint effect based on piecewise-far-field approximation in \cite{Cui_NF_wideband}. Moreover, the authors of \cite{Cui_NF_wideband_traning} developed an effective time-delay-based beam training scheme that capitalizes on the beam squint effect in the near-field communication, significantly enhancing the performance of near-field wideband systems. Unlike the aforementioned studies, which focus on modifying the BS hardware architecture, the authors of \cite{Myers_NF_wideband} proposed a low-complexity spatial coding technique to mitigate the misfocus effect in near-field wideband communication systems, offering an alternative solution with reduced hardware requirements.

\emph{2) RIS/STARS-Assisted Near-field Wideband Communications:} With the rapid advancement of RIS technology, RIS-assisted near-field wideband communication systems have emerged as a highly researched topic. However, the beam squint effect, resulting from frequency-independent beamforming at the RIS, also poses a significant challenge, limiting the system's performance. To tackle this challenge, substantial research efforts have been devoted across various fronts. In \cite{Cheng_RIS_NF_wideband}, the authors studied a RIS-assisted near-field wideband system and proposed a near-optimal phase shift design and a virtual-subarray-based phase shift design for single-user and multi-user scenarios to counter the beam squint effect, respectively. Similarly, in \cite{Wang_RIS_NF_wideband}, to counter the near-field double beam squint effect in RIS-assisted wideband communications, a TTD-based BS architecture and two specific RIS architectures, namely TTD-based RIS and virtual subarray-based RIS, were proposed to implement frequency-dependent active and passive beamforming, respectively. Further, the authors of \cite{Yang_RIS_NF_wideband} first characterized the near-field beam squint effect in both angle and distance domains for extremely large (XL)-RIS-assisted wideband mmWave systems and then formulated a two-dimensional (2-D) compressive sensing problem to recover the channel parameter using a spherical-domain dictionary based on the sparse nature of mmWave channels. In \cite{Hao_beam_squint}, the authors investigated beam squint issues in both far-field and near-field THz RIS communications and proposed a delay-adjustable metasurface RIS architecture to effectively reduce the beam gain loss caused by beam squint. While in \cite{Li_RIS_NF_wideband}, the authors explored joint channel and location sensing schemes for THz XL-RIS systems, where a frequency-selective polar-domain redundant dictionary is introduced to address the hybrid field beam squint effect. Following this, the authors of \cite{Li_RIS_wideband} analyzed RIS-based wideband integrated sensing and communication systems, where the beam squint effect on a uniform planar array RIS was first explored and then tackled by proposing an efficient beam scanning scheme with TTD units.

\subsection{Motivations and Contributions}
Given the above context, numerous studies have focused on addressing the beam squint effect in RIS-assisted near-field wideband communication systems, with various mitigation strategies being proposed. However, these approaches often depend on hardware-intensive architectures, such as TTD units for phase compensation, which significantly increase hardware requirements and deviate from the low-power design philosophy of the RIS. Recently, various flexible-antenna systems, such as fluid-antenna systems \cite{Wong_Fluid,Wong_Fluid_access} and movable antennas \cite{Zhu_movable,Ma_movable}, have been introduced into wireless communication systems. By attaching flexible cables to conventional antenna structures, the positions of antennas can be dynamically adjusted within a feasible region, providing additional degrees of freedom (DoFs) in the spatial domain. Therefore, by adjusting antenna positions, the geometric path delay can be modified to compensate for phase shifts across different frequencies, offering functionality similar to TTD units but with simpler and more energy-efficient mechanisms. Inspired by these advancements, we propose extending flexible-antenna technology to the STARS, introducing a novel movable-element STARS (ME-STARS) design to explore its effectiveness in mitigating the beam squint for near-field wideband communications. To the best of our knowledge, this work represents the first exploration of this topic.

In contrast to TTD units, which directly use precise delay components for wideband consistency, the tuning of movable elements in ME-STARS focuses on spatial geometry. Therefore, controlling the movement of these elements is crucial for improving performance. To fully harness the potential of ME-STARS, we propose practical movement modes for adjusting element positions and investigate the optimization problem associated with them in near-field wideband communications. The main contributions of this paper are summarized as follows:
\begin{itemize}
    \item \textcolor{black}{We introduce a novel ME-STARS architecture and propose four distinct implementation modes: region-based (RB), horizontal-based (HB), vertical-based (VB), and diagonal-based (DB) modes.}
	\item We propose an ME-STARS-assisted near-field wideband communication framework, where a STARS with movable elements is utilized to facilitate downlink communication between a multi-antenna BS and multiple single-antenna users. To mitigate the beam squint effect in the proposed system, we formulate an optimization problem aimed at maximizing the sum achievable data rate among users by jointly optimizing the BS precoding, ME-STARS beamforming, and the positions of movable elements.
	\item We propose a two-layer algorithm to solve the resulting intractable problem for each mode. In the inner layer, we apply the block coordinate descent (BCD) framework to decouple the BS precoding and the ME-STARS beamforming, solving their corresponding subproblem alternatively. While in the outer layer, we use a particle swarm optimization (PSO)-based method to search the optimal positions of the movable elements in each iteration.
	\item Our numerical results depict that 1) ME-STARSs can effectively suppress the beam squint effect in near-field wideband communications; 2) the flexibility of movement and a wide movement region help facilitate the elimination of the beam squint effect; and 3) the DB mode achieves the most reasonable balance between performance gain, hardware overhead, and algorithmic complexity.
\end{itemize}
\subsection{Organization and Notations}
The remainder of the paper is organized as follows: Section II introduces the investigated system model and the formulated optimization problem. In Section III, an effective algorithm is explored to solve the resulting intractable problem. Subsequently, Section IV provides numerical results along with detailed discussions. Finally, Section V concludes the paper.

\emph{Notations}: Scalars, vectors, and matrices are denoted by lower-case, bold lower-case letters, and bold upper-case letters, respectively. $(\cdot)^T$ denotes the transpose, while $(\cdot)^H$ denotes the conjugate transpose. $\mathrm{Tr}(\cdot)$ and $\mathrm{Rank}(\cdot)$ denote the trace and rank of the matrices, respectively. $\|\cdot\|$ and $\|\cdot\|_2$ denote the norm and spectral norm, respectively. $\mathrm{diag}(\cdot)$ denotes the diagonalization operation on vectors. $\mathbb{C}^{M \times N}$ denotes the space of $M \times N$ complex valued matrices. $\mathbf{I}^{M\times M}$ denotes the unit matrix of order $M$. $\mathcal{CN}(\mu, \sigma^2)$ represents the distribution of a circularly symmetrical complex Gaussian random variable with a mean of $\mu$ and a variance of $\sigma^2$. 
\section{System Model and Problem Formulation}
\begin{figure}
	\setlength{\abovecaptionskip}{0cm}   
	\setlength{\belowcaptionskip}{0cm}   
	\setlength{\textfloatsep}{7pt}
	\centering
	\includegraphics[width=3.6in]{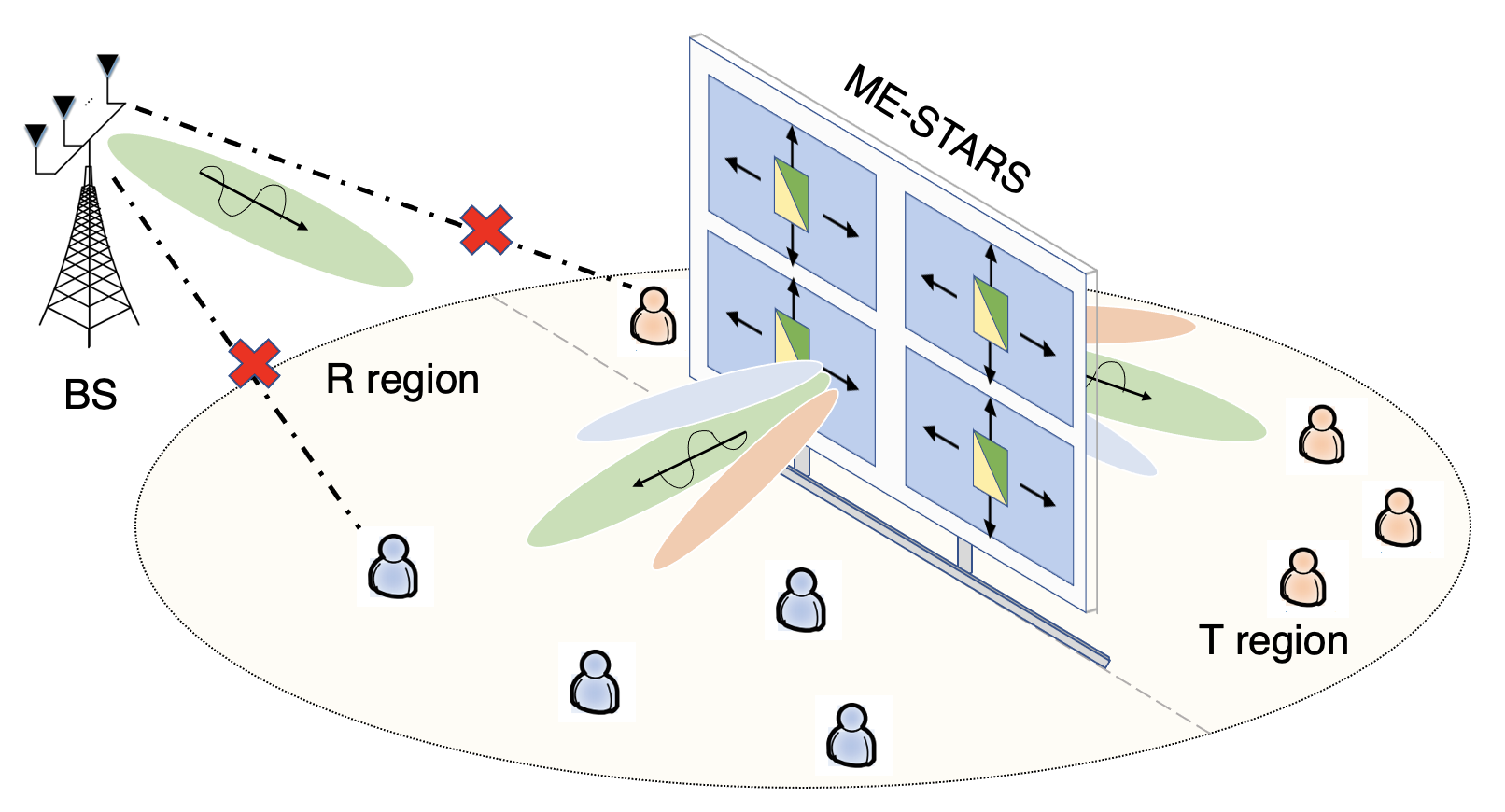}
	\caption{Illustration of an ME-STARS-assisted near-field wideband communication system.}
	\label{system}
\end{figure}
As shown in Fig. \ref{system}, we consider an ME-STARS-assisted THz wideband communication system, where a STARS with $M$ movable elements is deployed to enhance communication between an $N$-antenna BS and $K$ single-antenna users. Due to the severe path loss, the direct links between the BS and users in THz communication are more susceptible to obstructions and are therefore disregarded in this paper \cite{Yan_STAR_wideband}. Additionally, the high center frequency $f_c$ and considerable aperture size $D$ employed in this system allow the ME-STARS to achieve a sufficiently large Rayleigh distance, given by $\frac{2D^2}{\lambda_c}$, where $\lambda_c=\frac{c}{f_c}$ denotes the wavelength corresponding to the center frequency while $c$ is the speed of light \cite{Rayleigh_distance}. This allows us to assume that all users fall within its near-field communication region. Furthermore, we categorize the users into two regions: $K_T$ users are located in the transmission (T) region of the ME-STARS, forming the set $\mathcal{K_T}$, while the remaining $K_R$ users are located in the reflection (R) region, forming the set $\mathcal{K_R}$. These two sets satisfy $\mathcal{K_T} \cup \mathcal{K_R}=\mathcal{K}$ and $\mathcal{K_T} \cap \mathcal{K_R}= \varnothing$, where $\mathcal{K}$ represents the set of all users. Besides, the BS and the ME-STARS are typically separated by a significant distance, resulting in a far-field propagation scenario where signals exhibit plane wavefronts. 

\subsection{Wideband Channel Model}
For the wideband communication, the orthogonal frequency division multiplexing (OFDM) technique with $L$ subcarriers is applied to overcome the frequency-selective fading \cite{Wang_wideband}. Denoting $B$ as the bandwidth and $f_c$ as the central frequency, the frequency on the $l$-th subcarrier is given by $f_l=f_c+\frac{B}{L}(l-\frac{L+1}{2}), l=1,\cdots,L.$
Generally, the communication channel consists of a line-of-sight (LoS) path and a few non-line-of-sight (NLoS) paths. However, due to the property of high-frequency signals, the LoS path is usually dominant; hence, we assume that only the LoS path exists here.
Then, the channel coefficients from the BS to the ME-STARS and from the ME-STARS to user $k$ on subcarrier $l$ are denoted by $\mathbf{F}_l\in \mathbb{C}^{M\times N}$ and $\mathbf{h}_{l,k} \in \mathbb{C}^{M\times1}$. 
\textcolor{black}{Although the absence of active signal processing at the STARS, coupled with the element's dynamic nature, significantly complicates channel estimation}, these parameters in the channel response can be efficiently estimated using the segmentation method proposed in \cite{Wu_estimation} combined with the compressed sensing algorithm from \cite{Xiao_estimation}. Therefore, we assume that the channel state information (CSI) of all links is perfectly known at the BS. \textcolor{black}{It is important to note that the assumption of perfect CSI is made to establish an upper bound on performance, simplifying the analysis and providing fundamental insights. Similarly, assuming a pure LoS channel aligns with the typical use case of ME-STARS, where the goal is to create a dominant LoS path. These assumptions help streamline the analysis and emphasize key design principles, while future work will investigate the impact of practical challenges, such as NLoS components and CSI estimation errors.} 
%

\emph{1) BS-ME-STARS Far-Field Channel Model:} We adopt the commonly used far-field LoS model to fit the channel between them. In addition, we assume that the antennas are arranged in a ULA at the BS while the elements are arranged in a planar array (PA) at the ME-STARS. Let the coordinates of the BS be denoted by $\mathbf{b}=[x_B,y_B,z_B]^T$. The coordinates at the center of ME-STARS are denoted as $\mathbf{s}_c=[x_S, y_S, z_S]^T$, and the $m$-th movable element is coordinated by $\mathbf{s}_m=[x_m,y_m,z_m]^T$. Thus, the channel matrix of the $l$-th subcarrier between the BS and the ME-STARS is given by
\begin{align}
	\mathbf{F}_l=\sqrt{MN} \alpha_{0,l} \mathbf{a}_S\left(\varphi_A,\psi_A, f_l\right)\mathbf{a}_B^H\left(\varphi_D, f_l\right),
\end{align}
where $\alpha_{0,l}$ denotes the channel gain of the $l$-th subcarrier from the BS to the ME-STARS and is shown as
\begin{align}
  	\alpha_{0,l}=\frac{c}{4\pi f_l d_{B,S}}e^{-\frac{1}{2}K_{abs}(f_l) d_{B,S}},
\end{align} 
where $K_{abs}(f)$ and $d_{B,S}=\|\mathbf{b}-\mathbf{s}_c\|_2$ denote the medium absorption factor of frequency $f$ and distance from the BS to the ME-STARS, respectively. Besides, the $\varphi_A \in [0, 2\pi)$ and $\psi_A \in [-\frac{\pi}{2},\frac{\pi}{2}]$ denote the azimuth and elevation angle of arrival (AoA) associated with the ME-STARS, respectively, and $\varphi_D \in [-\frac{\pi}{2},\frac{\pi}{2}] $ denotes the angle of departure (AoD) of the BS. For the purpose of this analysis, we assume that the ME-STARS is deployed perpendicular to the $y$-axis within the constructed coordinate system. As a result, the corresponding array response vectors on frequency $f$ for the PA-type ME-STARS are denoted by $\mathbf{a_S}\left(\varphi_A,\psi_A, f\right)$, which is expressed as
\begin{align}
  	\mathbf{a}_S\left(\varphi_A,\psi_A, f\right)=&\frac{1}{\sqrt{M}}[e^{-j\frac{2\pi f}{c}\left(x_1\sin\varphi_A\sin\psi_A+z_1\cos\psi_A\right)}, \cdots, \nonumber \\
  	&e^{-j\frac{2\pi f}{c}\left(x_M\sin\varphi_A\sin\psi_A+z_M\cos\psi_A\right)}]^T.
\end{align}
Similarly, for the ULA-type BS with $N$ antennas, the array response vector $\mathbf{a}_B\left(\varphi_D, f\right)$ on frequency $f$ is given by 
\begin{align}
  	\mathbf{a}_B\left(\varphi_D, f\right)=\frac{1}{\sqrt{N}}[1,\cdots, e^{-j\frac{2\pi f d}{c}(N-1)\sin\varphi_D}]^T,
\end{align}
where $d$ denotes the antenna spacing of the BS and is set to half of the central wavelength, i.e., $d=\frac{\lambda_c}{2}$. 

\emph{2) ME-STARS-User Near-Field Channel Model:} For the channel from the ME-STARS to users, we adopt the near-field LoS channel model in the following discussion. Regarding the origin as the reference point, the position of the $k$-th user is given by $\mathbf{p}_{k}=[x_k,y_k,z_k]^T$. Furthermore, given positions of the ME-STARS $m$-th movable element $\mathbf{s}_{m}=[x_m,y_m,z_m]^T$, the normal near-field channel response vector on subcarrier $l$ is given by
\begin{align}
  	\mathbf{h}_{l,k}=\alpha_{l,k}\mathbf{a}_k(f_l), \forall k \in \mathcal{K},
\end{align}
where $\alpha_{l,k}=\frac{c}{4\pi f_l d_{S,k}}e^{-\frac{1}{2}K_{abs}(f_l) d_{S,k}}$ denotes the channel gain and $d_{S,k}=\|\mathbf{p}_{k}-\mathbf{s}_c\|_2$ denotes the distance between the $k$-th user and the ME-STARS. Furthermore, vector $\mathbf{a}_k(f_l)$ denotes the array reponse vector and is given by 
\begin{align}
	\mathbf{a}_k(f)=[e^{-j\frac{2\pi f}{c}\|\mathbf{p}_{k}-\mathbf{s}_1\|_2},\cdots, e^{-j\frac{2\pi f}{c}\|\mathbf{p}_{k}-\mathbf{s}_M\|_2}]^T.
\end{align}
\subsection{ME-STARS Model and Movement Modes}
\begin{figure}[]
	\subfigure[\textcolor{black}{RB mode}.]{\label{RB}
		\includegraphics[width= 1.65in]{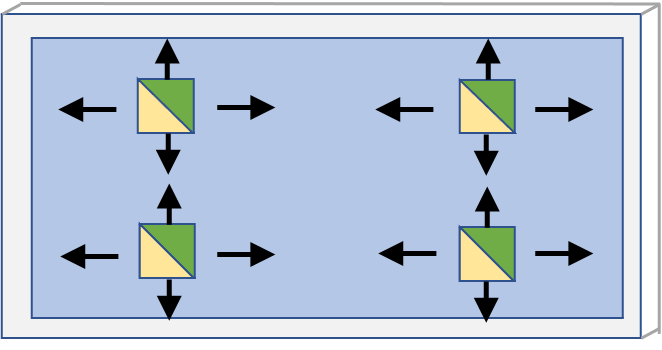}}
	\subfigure[HB mode.]{\label{HB}
		\includegraphics[width= 1.65in]{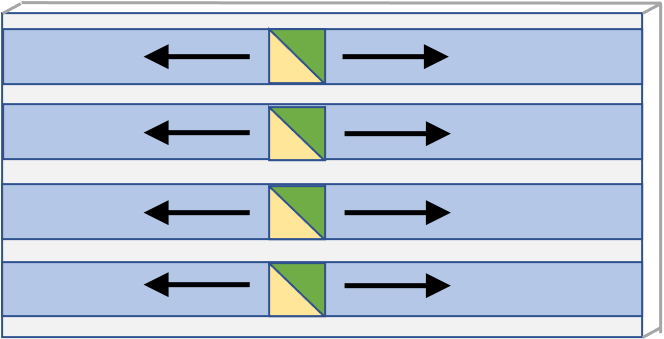}}
	\subfigure[VB mode.]{\label{VB}
		\includegraphics[width= 1.65in]{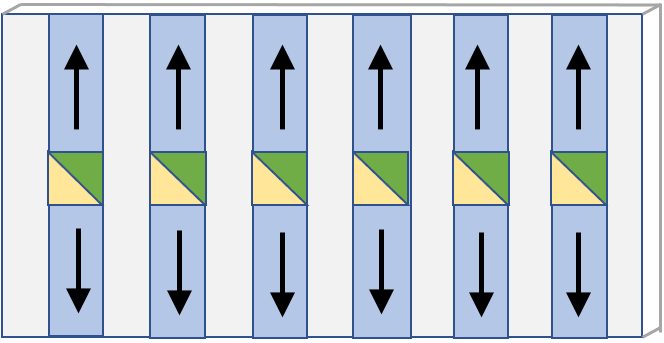}}
	\subfigure[DB mode.]{\label{DB}
		\includegraphics[width= 1.65in]{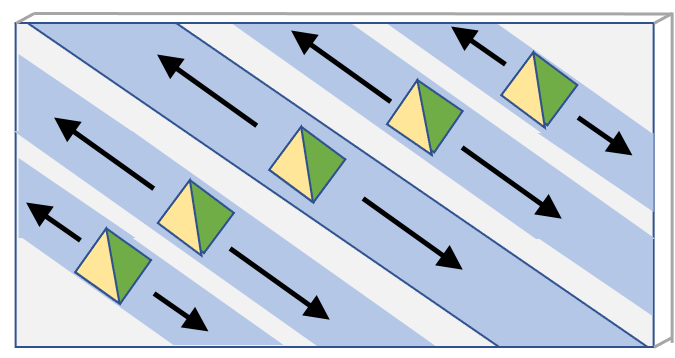}}
	\caption{\textcolor{black}{Illustration of proposed four practical ME-STARS modes.} } 
	\label{Movable-model}
\end{figure}
In this work, the ES protocol is operated in the ME-STARS. The amplitude adjustments of the $m$-th element for transmission and reflection are denoted as $\beta^t_m$ and $\beta^r_m$. Due to the law of conservation of energy, the constraints $\beta^t_m, \beta^r_m \in [0,1]$ and $\beta^t_m+\beta^r_m=1, \forall m \in\mathcal{M}=\{1,2,\cdots,M\}$ always hold. In addition, for each element of the ME-STARS, there are two independent phase shifts for transmission and reflection, which are denoted by $\theta^t_m$ and $\theta^r_m$, respectively. Similar to conventional RISs, we assume the design of the phase shifts is continuous, i.e., $\theta^t_m, \theta^r_m \in [0,2\pi), \forall m\in\mathcal{M}$. Thus, the transmission- and reflection-coefficient matrices of the ME-STARS can be given by $\mathbf{\Theta}_t=\textup{diag}\left(\sqrt{\beta_1^{t}}e^{j\theta^t_1},\cdots,\sqrt{\beta_M^{t}}e^{j\theta^t_M}\right)$ and $\mathbf{\Theta}_r=\textup{diag}\left(\sqrt{\beta_1^{r}}e^{j\theta^r_1},\cdots,\sqrt{\beta_M^{r}}e^{j\theta^r_M}\right)$, respectively. 

 Particularly, we further assume that the ME-STARS is a square array of $[-\frac{A}{2},\frac{A}{2}] \times [-\frac{A}{2},\frac{A}{2}]$, \textcolor{black}{where each element is connected to the central processing unit (CPU) via the flexible cable. By cooperatively performing the movement controllers like stepper motors or servo motors, an element can be moved in a 2-D area by simultaneously adjusting its $x$-axis and $z$-axis coordinates, or in a one-dimensional (1-D) area by modifying either the $x$-axis or $z$-axis coordinate individually in this plane}. Inspired by this, we propose four practical movement modes to maximize the potential of ME-STARS, as illustrated in Fig. \ref{Movable-model}.

1) \emph{Region-based Movement Mode}: For RB, as shown in Fig. \ref{RB}, each element of the STARS is allowed to move freely in all directions within a 2-D area, i.e, $x_m\in[-\frac{A}{2},\frac{A}{2}]$ and $z_m\in[-\frac{A}{2},\frac{A}{2}], \forall m\in\mathcal{M}$. This mode significantly enhances design flexibility but \textcolor{black}{typically requires the coordinated operation of multiple motors, introducing additional challenges for practical implementation.}

2) \emph{Horizontal-based Movement Mode}: For HB, as shown in Fig. \ref{HB}, each STARS element is constrained to move only horizontally within its row to improve communication quality. In this case, the $z$-axis coordinate of each element is fixed, while its $x$-axis coordinate is restricted to $[-\frac{A}{2},\frac{A}{2}]$. \textcolor{black}{More specifically, the fixed $z$-axis coordinates can be defined as $z_m=z_S+(Z_m-\frac{Z_{num}+1}{2})\delta_H$, where $Z_{num}$ represents the total number of the slides, $Z_m$ denotes the slide index on which the $m$-th element is positioned, and $\delta_H$ denotes the spacing between neighboring slides.} While this mode limits design freedom, it simplifies implementation. Typically, it can be achieved by adding a sliding track to each row of the existing STARS structure.

3) \emph{Vertical-based Movement Mode}: For VB, as shown in Fig. \ref{VB}, each STARS element is restricted to movement in a 1-D area similar to the HB mode. However, in this mode, the elements move exclusively in the vertical direction. As a result, the $x$-axis coordinate of each element is fixed, while its $z$-axis coordinate is restricted to $[-\frac{A}{2},\frac{A}{2}]$. \textcolor{black}{Similarly to the HB mode, the coordinate $x_m$ of the $m$-th element is defined as $x_m=x_S+(X_m-\frac{X_{num}+1}{2})\delta_V$, where $X_{num}$ denotes the total number of the slides, $X_m$ denotes the slide index corresponding to the position of the $m$-th element, and $\delta_V$ denotes the spacing between neighboring slides.}

4) \emph{Diagonal-based Movement Mode}: For DB, as shown in Fig. \ref{DB}, each STARS element is enabled to move along the diagonal direction to overcome the coverage deficiencies of 1-D movement only. In addition, compared to RB, it can still adopt a track-embedded hardware design idea similar to HB and VB modes. Therefore, this mode strikes a good trade-off between mobility flexibility and implementation difficulty. In particular, the position of each element in this mode satisfies $x_m\in[-\frac{A}{2},\frac{A}{2}]$ and $z_m\in[-\frac{A}{2},\frac{A}{2}]$. However, since the movement is limited to a diagonal shift, $x_m$ and $z_m$ must satisfy additional constraints, such as $z_m=x_m+\delta_D$, where $\delta_D$ is a constant and is determined by the constructed coordinate system.
\subsection{Signal Transmission Model}
Due to its small number of antennas, we adopt a fully digital baseband precoder at the BS, where a dedicated beam, denoted by $\mathbf{w}_{l,k}\in\mathbb{C}^{N\times 1}$, is assigned to user $k$ on subcarrier $l$. Accordingly, the transmitted signal at the BS on the $l$-th subcarrier is expressed as
\begin{align}
	\mathbf{x}_l=\sum_{k=1}^{K}\mathbf{w}_{l,k}x_{l,k},
\end{align}
where $x_{l,k}$ denotes the information-bearing signal for user $k$ on subcarrier $l$ and satisfies with $\mathbb{E}\left[|x_{l,k}|^2\right]=1$. Assuming that all beams are independent of one another, the total transmit power at the BS can be expressed as follows:
\begin{align}
	P_T=\sum_{l=1}^LP_l=\sum_{l=1}^{L}\mathbb{E}\{\mathbf{x}_l^H\mathbf{x}_l\}=\sum_{l=1}^{L}\sum_{k=1}^{K}\mathbf{w}^H_{l,k}\mathbf{w}_{l,k}.
\end{align}
In particular, constrained by the limited available energy, the total transmit power cannot exceed the maximum allowable power budget $P_{\max}$, i.e., $P_T \leq P_{\max}$. For the purpose of analysis, we further assume that the power allocated by the BS to each subcarrier is identical \cite{wideband_power}, i.e., $P_l\leq \frac{P_{\max}}{L}.$

Due to obstructions blocking the direct links between the BS and the users, the signals transmitted by the BS must rely on the ME-STARS for relay to reach the users. Consequently, the signal received by user $k$ on the $l$-th subcarrier can be expressed as follows:
\begin{align}
y_{l,k}=\mathbf{h}^H_{l,k}\mathbf{\Theta}_{s_k}\mathbf{F}_l\mathbf{x}_l+n_{l,k},
\end{align}
where $s_k \in \{t,r\}$ indicates that user $k$ is located in the T region or R region. Besides, $n_{l,k} \sim \mathcal{CN}(0, \sigma^2)$ denotes the additive white Gaussian noise at user $k$ on the $l$-th subcarrier. As a consequence, the signal-to-interference-plus-noise ratio (SINR) for user $k$ on subcarrier $l$ is expressed as 
\begin{align}
	\textup{SINR}_{l,k}=\frac{\big|\mathbf{u}^H_{s_k}\mathbf{H}_{l,k}\mathbf{w}_{l,k}\big|^2}{\sum^{K}_{i=1,i\ne k}\big|\mathbf{u}^H_{s_k}\mathbf{H}_{l,k}\mathbf{w}_{l,i}\big|^2+\sigma^2},
\end{align}
where $\mathbf{u}_{s_k}=[\sqrt{\beta^{s_k}_1}e^{\theta^{s_k}_1},\cdots, \sqrt{\beta^{s_k}_M}e^{\theta^{s_k}_M}]^T, s_k \in \{t,r\}$ represents the ME-STARS tuning vector. $\mathbf{H}_{l,k}= \textup{diag}(\mathbf{h}_{l,k})\mathbf{F}_{l} \in \mathbb{C}^{M\times N}$ denotes the cascaded channel from the BS to user $k$ on subcarrier $l$. Therefore, the achievable date rate in bit/s/Hz of user $k$ is given by
\begin{align}
	R_k=\sum_{l=1}^{L}\textup{log}_2(1+\textup{SINR}_{l,k}).
\end{align}
\subsection{Beam Squint Mitigation Via Element Position Adjustment}
\begin{figure}
	\setlength{\abovecaptionskip}{0cm}   
	\setlength{\belowcaptionskip}{0cm}   
	\setlength{\textfloatsep}{7pt}
	\centering
	\includegraphics[width=3.6in]{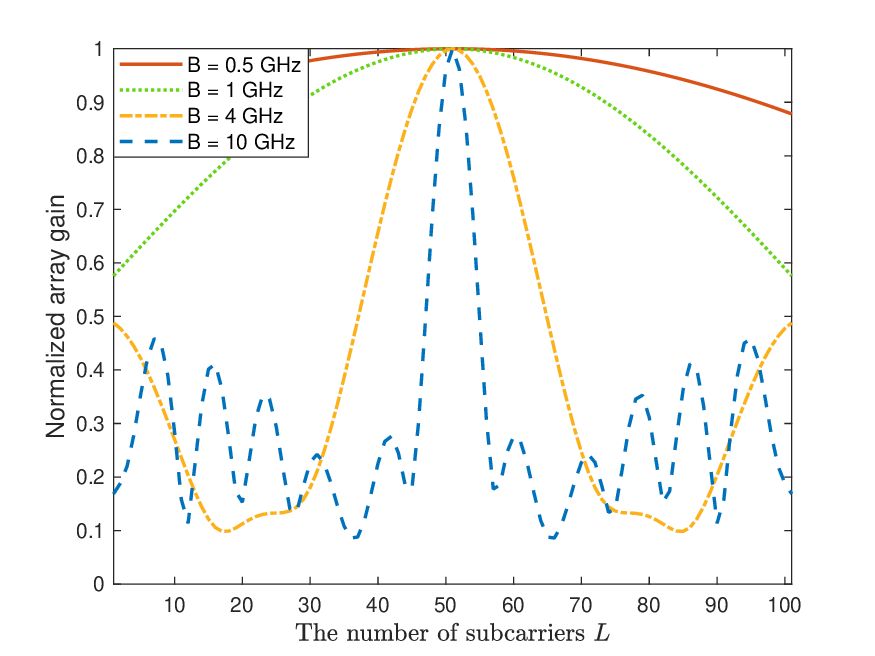}
	\caption{Normalized array gain versus different bandwidths.}
	\label{beam_squint}
\end{figure}
From the above model, it is evident that the far-field channel between the BS and the ME-STARS, the near-field channel between the ME-STARS and users, and the digital precoding vectors at the BS are all frequency-dependent. In contrast, the phase adjustment matrices at the ME-STARS remain frequency-independent. 
As a result, unlike \cite{Yan_STAR_wideband}, which highlights beam squint effects across three components, this paper focuses on the two dominant beam squint effects specifically associated with the ME-STARS.
Following this, the reflection/transmission normalized array gain for user $k$ on the $l$-th subcarrier, denoted as $g_k(f_l)$, is computed as \eqref{array_gain}, as shown at the top of the next page,
\begin{figure*}[t]
	\normalsize
	\begin{align}\label{array_gain}
		g_k(f_l)=\frac{1}{M}\big|\mathbf{a}^H_{k}(f_l)\mathbf{\Theta}_{s_k}\mathbf{a}_S(\varphi_A,\psi_A,f_l)\big| 
		=\frac{1}{M}\bigg|\sum_{m=1}^{M}e^{-j\frac{2\pi f_l}{c}\big[\|\mathbf{p}_k-\mathbf{s}_m\|_2+x_m\sin\varphi_A\sin\psi_A+z_m\cos\psi_A\big]}e^{j\theta_m^{s_k}}\bigg|
	\end{align}
	\hrulefill\vspace*{0pt}
\end{figure*}
where $\theta_m^{s_k}$ denotes the adjustable phase shifts of the $m$-th ME-STARS element and its corresponding amplitude $\beta_m^{s_k}$ is set to an arbitrary and reasonable constant in this subsection. Accordingly, when the ME-STARS phase configuration is designed based on the center frequency $f_c$, i.e., 
\begin{align}
	\theta_m^{s_k}=e^{j\frac{2\pi f_c}{c}\big[\|\mathbf{p}_k-\mathbf{s}_m\|_2+x_m\sin\varphi_A\sin\psi_A+z_m\cos\psi_A\big]}.
\end{align}
The normalized array gain is further computed as
\begin{align}\label{normalized}
	g_k(f_l)\!\!=\!\!\frac{1}{M}\bigg|\!\sum_{m=1}^{M}\!e^{j\frac{2\pi (f_c\!-\!f_l)}{c}\!\big[\|\mathbf{p}_k\!-\mathbf{s}_m\|_2+x_m\!\sin\!\varphi_A\!\sin\!\psi_A\!+z_m\!\cos\!\psi_A\big]}\!.
\end{align}
As can be observed, the generated beam achieves a maximum array gain of 1 only when $f_l=f_c$, indicating that it is perfectly aligned with the target location on this specific subcarrier. However, for other subcarrier frequencies, this alignment is not maintained, causing the signal reflected or transmitted by the ME-STARS to focus on unintended locations. This misalignment results in the beam squint effect, where the array gain falls below 1 for frequencies other than $f_c$.
To better illustrate this effect, we plot the normalized array gain for a specific user against different bandwidths $B$ in Fig. \ref{beam_squint}, where we set $f_c$ = 40 GHz and $L$ = 101. The results show that as the bandwidth increases, the beam squint effect becomes increasingly severe.

To overcome this issue and maximize $g_k(f_l)$ across all frequencies, it is crucial to eliminate the frequency-dependent terms in the exponent. Mathematically, this requires satisfying the following equation for all subcarriers:
\begin{align}\label{Position}
	\|\mathbf{p}_k-\mathbf{s}_m\|_2+x_m\sin\varphi_A\sin\psi_A+z_m\cos\psi_A=0.
\end{align}

\textcolor{black}{In conventional STARS with fixed element positions, satisfying this equation is challenging. Therefore, additional costly TTD structures are employed for precise delay compensation \cite{Yan_STAR_wideband}.
However, with the introduction of mobility to ME-STARS elements, spatial delay can be smartly adjusted through flexible positional movement to satisfy the equation. For a more detailed analysis, it is important to examine the terms on the left-hand side of \eqref{Position}. Among them, the first term, $\|\mathbf{p}_k-\mathbf{s}_m\|_2$, represents the distance between the ME-STARS element and the user, which is always non-negative. In contrast, the remaining terms, $x_m\sin\varphi_A\sin\psi_A+z_m\cos\psi_A$, capture the signal propagation from the BS to the ME-STARS and can take positive or negative values depending on the angles $\varphi_A$ and $\psi_A$. Therefore, by strategically adjusting the positions of the ME-STARS elements, their distances to the user can be modified to counterbalance the contributions of the angular terms, i.e., $\|\mathbf{p}_k-\mathbf{s}_m\|_2$ can be tuned to exactly offset $x_m\sin\varphi_A\sin\psi_A+z_m\cos\psi_A$. This enables the equation \eqref{Position} to be satisfied, ensuring that the signal focusing remains accurate across frequencies. Thus, the mobility of ME-STARS elements provides a dynamic and effective way to mitigate the beam squint effect by spatially compensating for frequency-dependent phase shifts.}

\subsection{Problem Formulation}
In this paper, we formulate a problem to maximize the sum achievable rate across users, where 
the active BS precoding vectors $\{\mathbf{w}_{l,k}\}$, the passive ME-STARS beamforming vectors $\{\mathbf{u}_s\}$, and the element position vectors $\{\mathbf{s}_m\}$ are jointly optimized to achieve this objective. The problem is defined as follows:
\begin{subequations}\label{Problem1}
\begin{align}
	&\label{P1_C0}\  \max_{\{\mathbf{w}_{l,k}\},\{\mathbf{u}_s\},\{\mathbf{s}_m\}}  \sum^K_{k=1}\ R_k\\ 
	&\label{P1_C1} {\rm s.t.} \ \sum^K_{k=1} \mathbf{w}^H_{l,k}\mathbf{w}_{l,k} \leq P_{\max}/L,\  l=1,\cdots,L,\\
	&\label{P1_C2} \quad \ \ \mathbf{s}_m \in \mathcal{D}, \forall m \in \mathcal{M},\\
	&\label{P1_C3} \quad \ \ \|\mathbf{s}_m-\mathbf{s}_o\|_2 \geq d_{\min}, m,o \in \mathcal{M}, m \ne o,\\ 
	&\label{P1_C4} \quad \ \  \mathbf{u}_s\!=\![\sqrt{\beta_1^{s}}e^{j\theta^{s}_1},\cdots\!,\sqrt{\beta_M^s}e^{j\theta^{s}_M}]^T, s \in \{t,r\}, \\
	&\label{P1_C5} \quad \ \ \theta^{s}_m \in [0,2\pi),\forall m \in \mathcal{M}, s\in\{t,r\},\\
	&\label{P1_C6} \quad\ \ \beta_m^{t}, \beta_m^{r}\in [0,1],\beta_{m}^{t}+\beta_{m}^{r}=1, \forall m\in \mathcal{M},
\end{align}
\end{subequations}
where constraint $\eqref{P1_C1}$ denotes the total transmit power budget at the BS. $\mathcal{D}$ in constraint $\eqref{P1_C2}$ represents the feasible movable area for each ME-STARS element, and $d_{\min}$ in constraint $\eqref{P1_C3}$ represents the minimum distance to prevent coupling effects among movable elements. More particularly, for each element, $\mathcal{D}$ represents a 1-D area in HB and VB modes, whereas it represents a 2-D area in RB and DB modes. In addition, $\eqref{P1_C5}$ and $\eqref{P1_C6}$ are the constraints on phase shift and amplitude for ME-STARS elements, respectively.

As can be observed, all the optimization variables are presented in a highly-coupled form within the objective function. Furthermore, unlike the fixed-position element-enabled STARS, the introduction of movable elements brings additional spatial movement constraints that must be accounted for. Moreover, the movement of these elements influences both the channels between the BS and the ME-STARS, as well as between the ME-STARS and users, rendering the position design significantly more complicated than in systems with general movable antennas at the BS. In summary, these factors create a highly challenging problem \eqref{Problem1} that existing methods cannot adequately solve. To overcome this issue, we propose an efficient algorithm in the next section.

\section{Proposed Solutions}
In this section, we propose a two-layer algorithm to solve the resulting intractable problem. In the outer layer, the element position vectors $\{\mathbf{s}_m\}$ are updated via a PSO-based search method in an iterative manner. While in the inner layer with fixed element positions, the remaining highly-coupled variables are first decomposed into two blocks, i.e., $\{\mathbf{w}_{l,k}\}$ and $\{\mathbf{u}_s\}$, and then the BCD method is employed to solve their corresponding subproblems alternatively.
\subsection{Inner Layer: BCD Method to Optimize BS Precoding and ME-STARS Beamforming }
First, we focus on the inner layer optimization problem. With given element position vectors $\{\mathbf{s}_m\}$, the original problem is reduced into a joint BS precoding and ME-STARS beamforming optimization problem as
\begin{subequations}\label{P2}
	\begin{align}
		&\label{P2_C0}\  \max_{\{\mathbf{w}_{l,k}\},\{\mathbf{u}_s\}} \sum^K_{k=1} R_k \\ 
		&\label{P2_C1} \quad \quad {\rm s.t.}\quad \eqref{P1_C1}, \eqref{P1_C4}, \eqref{P1_C5}, \eqref{P1_C6}.
	\end{align}
\end{subequations}
In particular, problems similar to \eqref{P2} have been extensively investigated in narrow-band systems, where advanced optimization algorithms such as penalty-based multi-layer joint beamforming optimization were developed \cite{Zhu_SWIPT}. However, these algorithms often entail high computational complexity, and thus applying them directly to our formulated problem within the outer PSO loop would substantially increase the overall complexity of the algorithm. To overcome this challenge, we propose a simpler BCD-based framework to alternately optimize the BS precoding and the ME-STARS beamforming. In this framework, the BS precoding is efficiently handled using the SDR method to obtain the optimal solution directly, while the complexity of ME-STARS beamforming design is reduced by addressing the rank-one constraints through SDR combined with the Gaussian randomization.

Next, we extend the optimization variables from the vector dimension to the matrix dimension, as defined by $\mathbf{W}_{l,k}\triangleq\mathbf{w}_{l,k}\mathbf{w}^H_{l,k} \in \mathbb{C}^{N\times N}$ and $\mathbf{U}_{s}\triangleq\mathbf{u}_s\mathbf{u}^{H}_s \in \mathbb{C}^{M\times M}$. To this end, new optimal variables $\{\mathbf{W}_{l,k}\}$ and $\{\mathbf{U}_s\}$ are constrained to be semi-positive definite matrices. Accordingly, the original problem \eqref{P2} can be reformulated as a semi-positive definite program (SDP) problem in the following form \cite{Huang_SDP}:
\begin{subequations}\label{P3}
	\begin{align}
		&\label{P3_C0}\!\!\max_{\{\!\mathbf{W}_{l,k}\!\}\!,\{\!\mathbf{U}_s\!\}}\!\! \sum^K_{k=1}\!\sum^L_{l=1}\! \log_2\!\!\left(\!\!1\!+\!\frac{\mathrm{Tr}\left(\mathbf{W}_{l,k}\mathbf{H}^H_{l,k}\mathbf{U}_{s_k}\mathbf{H}_{l,k}\right)}{\sum^K_{i=1,i \ne k}\!\mathrm{Tr}\!\left(\!\mathbf{W}_{l,i}\!\mathbf{H}^H_{l,k}\!\mathbf{U}_{s_k}\!\mathbf{H}_{l,k}\!\right)\!\!+\!\sigma^2}\!\!\!\right) \\
		&\label{P3_C1} \quad \quad {\rm s.t.} \ \sum^K_{k=1} \mathrm{Tr} \left(\mathbf{W}_{l,k}\right) \leq P_{\max}/L, \ l=1,\cdots,L,\\
		&\label{P3_C2} \quad \quad \quad \quad \mathbf{W}_{l,k} \succeq 0, \forall k \in \mathcal{K}, l=1,\cdots,L,\\
		&\label{P3_C3} \quad \quad \quad \quad \mathrm{Rank}\left(\mathbf{W}_{l,k}\right)=1, \forall k \in \mathcal{K}, l=1,\cdots,L,\\
		&\label{P3_C4} \quad \quad \quad \quad \mathbf{U}_s \succeq 0, s \in \{t,r\},\\
		&\label{P3_C5} \quad \quad \quad \quad \mathrm{Rank}\left(\mathbf{U}_s\right)=1, s \in \{t,r\},\\
		&\label{P3_C6} \quad \quad \quad \quad [\mathbf{U}_t]_m+[\mathbf{U}_r]_m=1, \forall m\in \mathcal{M}.
	\end{align}
\end{subequations}
Subject to the complexity of the logarithmic operation and the non-convexity of the differential form in the objective function, problem \eqref{P3} is challenging to solve directly. To address this issue, we introduce two sets of auxiliary variables $\{C_{l,k}\}$ and $\{D_{l,k}\}$, with $\frac{1}{C_{l,k}} \leq \mathrm{Tr}\left(\mathbf{W}_{l,k}\mathbf{H}^H_{l,k}\mathbf{U}_{s_k}\mathbf{H}_{l,k}\right)$ and $D_{l,k} \geq \sum_{i=1,i\ne k}^{K} \mathrm{Tr}(\mathbf{W}_{l,i}\mathbf{H}^H_{l,k}\mathbf{U}_{s_k}\mathbf{H}_{l,k})+\sigma^2$, to simplify the formulation \cite{Zhu_FP}. In this case, the $R_k$ in the objective function can be transformed as $\sum_{l=1}^{L}\log_2\left(1+\frac{1}{C_{l,k}D_{l,k}}\right)$. Next, using the first Taylor expansion of $\log_2(1+\frac{1}{C_{l,k}D_{l,k}})$ with respect to $C_{l,k}$ and $D_{l,k}$, we can obtain its convex lower boundary as 
\begin{align}
	&\sum_{l=1}^{L}\log_2\left(1+\frac{1}{C_{l,k}D_{l,k}}\right) \geq \sum_{l=1}^{L}\Bigg(\log_2\left(1+\frac{1}{C^{(j)}_{l,k}D^{(j)}_{l,k}}\right) \nonumber \\ 
	&-\frac{\log_2(e)\left(C_{l,k}-C^{(j)}_{l,k}\right)}{C^{(j)}_{l,k}\left(1+C^{(j)}_{l,k}D^{(j)}_{l,k}\right)}-\frac{\log_2(e)\left(D_{l,k}-D^{(j)}_{l,k}\right)}{D^{(j)}_{l,k}\left(1+C^{(j)}_{l,k}D^{(j)}_{l,k}\right)} \Bigg)\! \triangleq\! \widetilde{R}^{(j)}_k, 
\end{align}
where $C^{(j)}_{l,k}$ and $D^{(j)}_{l,k}$ denote the local points of $C_{l,k}$ and $D_{l,k}$ in the $j$-th iteration, respectively. By applying the above transformations, problem \eqref{P3} can be converted into 
\begin{subequations}\label{SDP}
	\begin{align}
		&\label{P4_0}\max_{\{\mathbf{W}_{l,k}\},\{\mathbf{U}_s\},\{C_{l,k}\},\{D_{l,k}\}} \sum^K_{k=1} \widetilde{R}^{(j)}_k \\
		&\label{P4_1} \quad \quad {\rm s.t.} \ \ C_{l,k} \geq 0, D_{l,k} \geq 0, \\
		& \quad \quad \quad \quad  \frac{1}{C_{l,k}}\leq \mathrm{Tr}\left(\mathbf{W}_{l,k}\mathbf{H}^H_{l,k}\mathbf{U}_{s_k}\mathbf{H}_{l,k}\right), \forall k \in \mathcal{K}, \nonumber \\
		&\label{P4_2} \quad \quad \quad \quad l =1,\cdots,L, \\
		&\quad \quad \quad \quad D_{l,k} \geq \! \! \sum_{i=1,i\ne k}^{K} \! \! \mathrm{Tr}(\mathbf{W}_{l,i}\mathbf{H}^H_{l,k}\mathbf{U}_{s_k}\mathbf{H}_{l,k})+\sigma^2, \forall k \in \mathcal{K}, \nonumber \\
		&\label{P4_3} \quad \quad \quad \quad l =1,\cdots,L, \\
		&\label{P4_4} \quad \quad \quad \quad \eqref{P3_C1}-\eqref{P3_C6}.
	\end{align}
\end{subequations}

Now, the main challenge in solving problem \eqref{SDP} lies in the tight coupling of the optimization variables. To circumvent this, we employ the BCD framework to decouple the BS precoding matrices 
$\{\mathbf{W}_{l,k}\}$ and the ME-STARS beamforming matrices $\{\mathbf{U}_s\}$. Subsequently, this approach enables us to solve their corresponding subproblems in an iterative manner.

\emph{1) BS Precoding Optimization with SDR:} With the given ME-STARS beamforming matrices $\{\mathbf{U}_s\}$, the non-convex rank-one constraint \eqref{P3_C3} poses the major obstacle in solving this subproblem. To address this, the SDR method can be employed to relax this challenging constraint, allowing us to solve a simplified and more tractable problem as
\begin{subequations}\label{BS_Precoding}
	\begin{align}
		&\label{P5_C0}\max_{\{\mathbf{W}_{l,k}\},\{C_{l,k}\},\{D_{l,k}\}} \widetilde{R}^{(j)}_k \\
		&\label{P5_C1} \quad \quad {\rm s.t.} \eqref{P3_C1}, \eqref{P3_C2}, \eqref{P4_1}, \eqref{P4_2}, \eqref{P4_3}.
	\end{align}
\end{subequations}
This is a standard SDP problem, which can be efficiently solved using existing solvers, such as CVX \cite{cvx}. Regarding the rank-one constraint, \cite[\textbf{Theorem 1}]{Rank-one} provides a theoretical proof demonstrating that $\mathrm{Rank}\left(\mathbf{W}_{l,k}\right) \leq 1$ always holds for each user and subcarrier. Therefore, the solutions to the relaxed problem \eqref{SDP} always satisfy the rank-one constraint \eqref{P3_C3} when $\mathbf{w}_{l,k} \ne 0$, implying that these solutions are also optimal solutions to the original problem. As a result, we can solve the BS precoding optimization subproblem directly by exploiting the SDR method. On this basis, the optimal BS precoding vectors $\{\mathbf{w}_{l,k}\}$ can be obtained through Cholesky decomposition, e.g., $\mathbf{W}^*_{l,k}=\mathbf{w}^*_{l,k}(\mathbf{w}^*_{l,k})^H$.

\emph{2) ME-STARS Beamforming Optimization with SDR and Gaussian Randomization:} Next, with the given BS precoding matrices $\{\mathbf{W}_{l,k}\}$, the remaining optimization variables are ME-STARS beamforming matrices $\{\mathbf{U}_s\}$, along with two sets of auxiliary variables $\{C_{l,k}\}$ and $\{D_{l,k}\}$. In this case, the original problem \eqref{SDP} is simplified as
\begin{subequations}\label{STAR_beamforming}
	\begin{align}
		&\label{P6_C0}\max_{\{\mathbf{U}_{s}\},\{C_{l,k}\},\{D_{l,k}\}} \widetilde{R}^{(j)}_k \\
		&\label{P6_C1} \quad \quad {\rm s.t.} \eqref{P3_C4}, \eqref{P3_C5}, \eqref{P3_C6}, \eqref{P4_1}, \eqref{P4_2}, \eqref{P4_3}.
	\end{align}
\end{subequations}
The difficulty in designing ME-STARS beamforming arises from the non-convexity of the rank-one constraint in \eqref{P3_C5}. While the SDR method can be applied to relax the rank-one constraint, transforming the problem into a standard SDP that can be directly solved by the CVX solver, the resulting solutions do not strictly satisfy the rank-one constraint in \eqref{P3_C5} due to the relaxation. To address this issue, the Gaussian randomization method can be employed to reconstruct the solutions, as suggested in \cite{Wu_Gaussian}. The process is as follows: we first use the singular value decomposition to decompose the solution $\{\mathbf{U}_s\}$ of the problem \eqref{STAR_beamforming} without considering  \eqref{P3_C5} as
\begin{align}
	\mathbf{U}_s\!=\!\mathbf{X}_s\mathbf{\Sigma}_s\mathbf{X}_s^H\!=\!\left(\mathbf{X}_s\sqrt{\mathbf{\Sigma}_s}\right)\left(\mathbf{X}_s\sqrt{\mathbf{\Sigma}_s}\right)^H\!\!,s \in \{t,r\},
\end{align}
where $\mathbf{X}_s$ is the unitary matrix, and $\mathbf{\Sigma}_s$ is the diagonal matrix. We assume $\mathbf{x}_s=\mathbf{X}_s\sqrt{\mathbf{\Sigma}_s}\mathbf{r}_s$, where $\mathbf{r}_s \in \mathcal{CN}(0,\mathbf{I}_{M})$. Thus, the near-optimal solution of $\mathbf{U}_s$ can be expressed as $\widetilde{\mathbf{U}}_s=\mathbf{x}_s(\mathbf{x}_s)^H$. Then, we can select the optimal value in $\widetilde{\mathbf{U}}_s$ to maximize the optimization goal, where $\mathbf{x}_s$ should be satisfied with all constraint of the original problem. Then, the near-optimal solution $\mathbf{x}_s$ is expressed as 
\begin{align}\label{Gaussian randomization}
	\mathbf{u}_s=e^{j arg\left([{\mathbf{x}_s}]_{(1:M)}\right)}.
\end{align}

It can be seen that in each iteration of the inner algorithm, the objective function is designed to be a non-decreasing value by alternately optimizing the two variable blocks, $\{\mathbf{w}_{l,k}\}$ and $\{\mathbf{u}_s\}$. In addition, there exists a theoretically achievable data rate for each user due to the channel capacity limitations. Together, these factors ensure that our proposed algorithm will converge to a stable value after several iterations. To clarify this process further, the specific details are summarized in \textbf{Algorithm \ref{alg:A}}.
\begin{algorithm}[!t]\label{method1}
	\caption{Proposed Inner Layer BCD-based Algorithm to Solve Problem \eqref{P2}.}
	\label{alg:A}
	\begin{algorithmic}[1]
		\STATE {Initialize the variables $\{\mathbf{u}^{(0)}_s\}$, convergence criterion $\varepsilon_0$, and the number of the Gaussian randomizations $G$.}
		\STATE Set the iteration index $j=0$ for the loop.
		\REPEAT 
		\STATE {Given ME-STARS beamforming $\{\mathbf{u}^{(j)}_s\}$, determine the optimal BS precoding vector $\{\mathbf{w}^{(j+1)}_{l,k}\}$ by solving problem \eqref{BS_Precoding}}.
		\STATE {Given BS precoding $\{\mathbf{w}^{(j+1)}_{l,k}\}$, solve problem \eqref{STAR_beamforming}, and recontruct the solution with \eqref{Gaussian randomization} to update$\{\mathbf{u}^{(j+1)}_s\}$}.
		\STATE Update $j \leftarrow j+1$.\\
		\UNTIL the iteration yield is below $\varepsilon_0$.\\
		\STATE {Output} the optimal solutions
		$\mathbf{u}^*_s=\mathbf{u}^{(j)}_s$ and $\mathbf{w}^*_{l,k}=\mathbf{w}^{(j)}_{l,k}.$
	\end{algorithmic}
\end{algorithm}
\subsection{Outer Layer: PSO-based Method to Optimize Element Positions}
In the outer layer, for the given $\{\mathbf{w}_{l,k}\}$ and $\{\mathbf{u}_s\}$, the original problem is reduced to an element position vectors $\{\mathbf{s}_m\}$ optimization problem as
\begin{subequations}\label{positions_optimization}
	\begin{align}
		&\label{P7_C0}\  \max_{\{\mathbf{s}_m\}}\ \ \sum^K_{k=1} R_k\\ 
		&\label{P&_C1} {\rm s.t.}\ \eqref{P1_C2}, \eqref{P1_C3}.
	\end{align}
\end{subequations}
However, given that the position of each element directly influences both the far-field channel from the BS to the ME-STARS and the near-field channel from the ME-STARS to users, optimizing $\{\mathbf{s}_m\}$ becomes an NP-hard problem that is challenging to solve using traditional mathematical methods. To this end, PSO, as a heuristic algorithm, presents itself as a viable approach to addressing this problem.

Firstly, for different movement modes of the ME-STARS, we define $I$ initialization particles corresponding to the position of each element as follows:
\begin{align}
	\widetilde{\mathbf{S}}^{(0)}_i\!\!=\!\! \begin{cases}\!
		[x_{i,1}^{(0)},x_{i,2}^{(0)},\cdots,x_{i,M}^{(0)}]^T, \textup{HB mode},\\ 
		[z_{i,1}^{(0)},z_{i,2}^{(0)},\cdots,z_{i,M}^{(0)}]^T, \textup{VB mode},\\
		[x_{i,1}^{(0)},z_{i,1}^{(0)},\cdots,x_{i,M}^{(0)},z_{i,M}^{(0)}]^T\!\!,\textup{RB/DB modes},
\end{cases}
\end{align}
In particular, assuming that the ME-STARS is a square array of $[-\frac{A}{2}$,$\frac{A}{2}] \times[-\frac{A}{2}$,$\frac{A}{2}]$, then $x^{(0)}_{i,1}$ and $z^{(0)}_{i,1}\in [-\frac{A}{2},\frac{A}{2}]$ for $1 \leq i \leq I, 1 \leq m \leq M$ ensure the feasibility of element positions' initialization. Similarly, the initial velocity of these particle swarms can be expressed as
\begin{align}
	\mathbf{V}^{(0)}_i\!\!=\!\! \begin{cases}
		[v_{x,i,1}^{(0)},v_{x,i,2}^{(0)},\cdots,v_{x,i,M}^{(0)}]^T, \textup{HB mode},\\ 
		[v_{z,i,1}^{(0)},v_{z,i,2}^{(0)},\cdots,v_{z,i,M}^{(0)}]^T, \textup{VB mode},\\
		[v_{x,i,1}^{(0)},v_{z,i,1}^{(0)},\cdots,v_{x,i,M}^{(0)},v_{z,i,M}^{(0)}]^T\!\!,\textup{RB/DB modes}.
	\end{cases}
\end{align}

Next, let $\widetilde{\mathbf{S}}_{i,p}$ and $\widetilde{\mathbf{S}}_{g}$ denote the personal best position of the $i$-th particle and the global best position of all particles, respectively. Following the iterative rules of the PSO algorithm \cite{PSO}, these parameters are updated in each iteration as follows:
\begin{align}\label{velocity}
\mathbf{V}^{(t+1)}_{i}=\omega\mathbf{V}^{(t)}_i\!+\!c_1\alpha_1(\widetilde{\mathbf{S}}_{i,p}\!-\!\widetilde{\mathbf{S}}^{(t)}_i)\!+\!c_2\alpha_2(\widetilde{\mathbf{S}}_{g}\!-\!\widetilde{\mathbf{S}}^{(t)}_i),
\end{align}
\begin{align}\label{position}
	\widetilde{\mathbf{S}}^{(t+1)}_i=\widetilde{\mathbf{S}}^{(t)}_i+\mathbf{V}^{(t+1)}_i,
\end{align}
where $t$ denotes the number of iterations, $c_1$ and $c_2$ are the personal and global learning factors, respectively. Correspondingly, $\alpha_1$ and $\alpha_2$ are set to random values between 0 and 1 to enhance the generalization of the personal and global searches. Besides, $\omega \geq 0$ denotes the inertia weight of the particle search. It is important to note that a larger $\omega$ value enhances global search capability but weakens personal search capability, while a smaller $\omega$ value has the opposite effect, strengthening personal search but reducing global search effectiveness. Therefore, to achieve better performance, it is essential to adjust its value dynamically. However, to avoid excessive design computational complexity, we adopt a linearly decreasing inertia weight as $\omega=\omega_{\max}-\left(\omega_{\max}-\omega_{\min}\right)t/T$ \cite{Xiao_PSO}, where $\omega_{\max}$ and $\omega_{\min}$ are predetermined upper and lower bounds of $\omega$, respectively, and $T$ is the maximum iteration number. Additionally, to ensure the constraints \eqref{P1_C2} are met, the position of each element must be further adjusted after each iteration to remain within the boundaries of the movable area. In other words, when the updated position exceeds the boundary, it is constrained to the boundary value; otherwise, it remains within its designated region.

Besides, the fitness function of each particle is associated with the maximization of the sum achievable user rate and determined by solving the problem \eqref{P2} in the inner layer with \textbf{Algorithm \ref{alg:A}}. For convenience, we define it as $\mathcal{R}(\widetilde{\mathbf{S}}_m)$ in this subsection. Furthermore, in order to ensure constraint \eqref{P1_C3}, we introduce a penalty factor $\eta>0$ to the fitness function and update it as follows:
\begin{align} \label{Fitness}
	\mathcal{F}(\widetilde{\mathbf{S}}_i)=\mathcal{R}(\widetilde{\mathbf{S}}_i)-\eta\mathcal{P}(\widetilde{\mathbf{S}}_i).
\end{align}
where $\mathcal{P}(\widetilde{\mathbf{S}}_i)$ is a penalty function that counts the number of instances in which the current element positions violate the collision constraints. It can be expressed as
\begin{align}
	\mathcal{P}(\widetilde{\mathbf{S}}_i)=\frac{1}{2}\sum_{m=1}^{M}\sum_{o=1,o\ne m}^{M} \mathbb{I}(\|\widetilde{\mathbf{s}}_{i,m}-\widetilde{\mathbf{s}}_{i,o}\|<d_{\min}),
\end{align}
where $\mathbb{I}$ is an indicator function that takes the value of 1 when the condition in parentheses is true and 0 otherwise. $\widetilde{\mathbf{s}}_{i,m}=[x_{i,m},y_{i,m},z_{i,m}]^T$ denotes the $m$-th element's position in particle $i$.
In this case, the smaller the value of $\mathcal{P}(\widetilde{\mathbf{s}}_m)$, the closer the obtained solution is to the strict element positions constraint \eqref{P1_C3} in the original problem \eqref{positions_optimization}. This makes the setting of the penalty factor $\eta$ essential to achieving a high-quality solution. Specifically, if the $\eta$ is set too large, the focus of the maximization objective $\mathcal{F}$ shifts to minimizing the penalty term. This ensures strict positional constraints but may cause the main objective $\mathcal{R}$ to converge prematurely. Conversely, if the $\eta$ is too small, the focus of $\mathcal{F}$ remains on maximizing $\mathcal{R}$, potentially leading to violations of the positional constraints. To strike a balance, we set $\eta$ to be on the order of 100 times the magnitude of the objective function $\mathcal{R}$.

Through iterative updates, each particle refines its own best position based on both individual experience and shared insights from the swarm. The global best position across all particles is adjusted, with its fitness value either improving or remaining stable until convergence is reached. In this way, a suboptimal solution for the $\{\mathbf{s}_m\}$ can be achieved. The exact processes are summarized in \textbf{Algorithm \ref{alg:B}}.
\begin{algorithm}[!t]\label{method2}
	\caption{Proposed Outer Layer PSO-based Algorithm to Solve Problem \eqref{positions_optimization}.}
	\label{alg:B}
	\begin{algorithmic}[1]
		\STATE Initialize the $I$ particles with position $\widetilde{\mathbf{S}}^{(0)}_i$ and $\mathbf{V}^{(0)}_i$, the number of iterations $t=1$.
		\STATE Initialize the local best position $\widetilde{\mathbf{S}}_{i,p}=\widetilde{\mathbf{S}}^{(0)}_{i}$ for each particle. 
		\STATE Determine the initial global best position of the particle swarm with $\widetilde{\mathbf{S}}_{g} = \arg \max \{\mathcal{F}(\widetilde{\mathbf{S}}^{(0)}_{1}),\cdots,\mathcal{F}(\widetilde{\mathbf{S}}^{(0)}_{I})\}$.
		\REPEAT 
		\STATE Update the inertia with $\omega=\omega_{\max}-\left(\omega_{\max}-\omega_{\min}\right)t/T$.
		\REPEAT 
		\STATE Update the velocity and position of particle $i$ according to \eqref{velocity} and \eqref{position}, respectively.\\
		\STATE Evaluate the fitess function $\mathcal{F}$ value for particle $i$ using \textbf{Algorithm \ref{alg:A}}.
		\STATE \textbf{if} $\mathcal{F}(\widetilde{\mathbf{S}}^{(t)}_{i})> \mathcal{F}(\widetilde{\mathbf{S}}_{i,p})$ \textbf{then}
		\STATE  \quad Update $\widetilde{\mathbf{S}}_{i,p} \leftarrow \widetilde{\mathbf{S}}^{(t)}_{i}$.
		\STATE  \textbf{end if}
		\STATE \textbf{if} $\mathcal{F}(\widetilde{\mathbf{S}}^{(t)}_{i})> \mathcal{F}(\widetilde{\mathbf{S}}_{g})$ \textbf{then}
		\STATE  \quad Update $\widetilde{\mathbf{S}}_{g} \leftarrow \widetilde{\mathbf{S}}^{(t)}_{i}$.
		\STATE  \textbf{end if}
		\STATE $i \leftarrow i+1$.
		\UNTIL $i>I$.
		\STATE $t \leftarrow t+1$.
		\UNTIL $t>T$.\\
		\STATE {Output} the optimal solutions
	 $\mathbf{s}^*_m$.
	\end{algorithmic}
\end{algorithm}

\subsection{Computational Complexity Analysis}
For the overall algorithm to solve problem \eqref{Problem1}, the internal SDP optimization problem for both BS precoding matrices and ME-STARS beamforming matrices can be solved using the interior-point method \cite{SDR}. Specifically, the rank-one constraint in the former is relaxed directly via the SDR method, while the rank-one constraint in the latter is handled by both the SDR and Gaussian randomization methods. As a result, the computational complexity of \textbf{Algorithm \eqref{alg:A}} in the inner layer is given by $\mathcal{O}\left(J(KN^{3.5}+2M^{3.5}+G)\right)$, where $J$ represents the number of iterations and $G$ is the number of randomizations within each iteration. In the outer layer, the PSO-based method is determined by the swarm size $I$ and the maximum number of iterations $T$. To determine the optimal element position vectors in \textbf{Algorithm 2}, \textbf{Algorithm 1} needs to be repeated at most $IT$ times\textcolor{black}{\!\cite{PSO_convergence}}. Therefore, the overall computational complexity of the algorithm is $\mathcal{O}\left(IT(J(KN^{3.5}+2M^{3.5}+G))\right)$.
\section{Numerical Results}
In this section, we present an appropriate simulation setup and provide various numerical results based on it to demonstrate the effectiveness of the ME-STARS in suppressing the beam squint effect for near-field wideband communication systems.
\subsection{Simulation Setup}
\begin{figure}[t]
	\setlength{\abovecaptionskip}{0cm}   
	\setlength{\belowcaptionskip}{0cm}   
	\setlength{\textfloatsep}{7pt}
	\centering
	\includegraphics[width=3.0in]{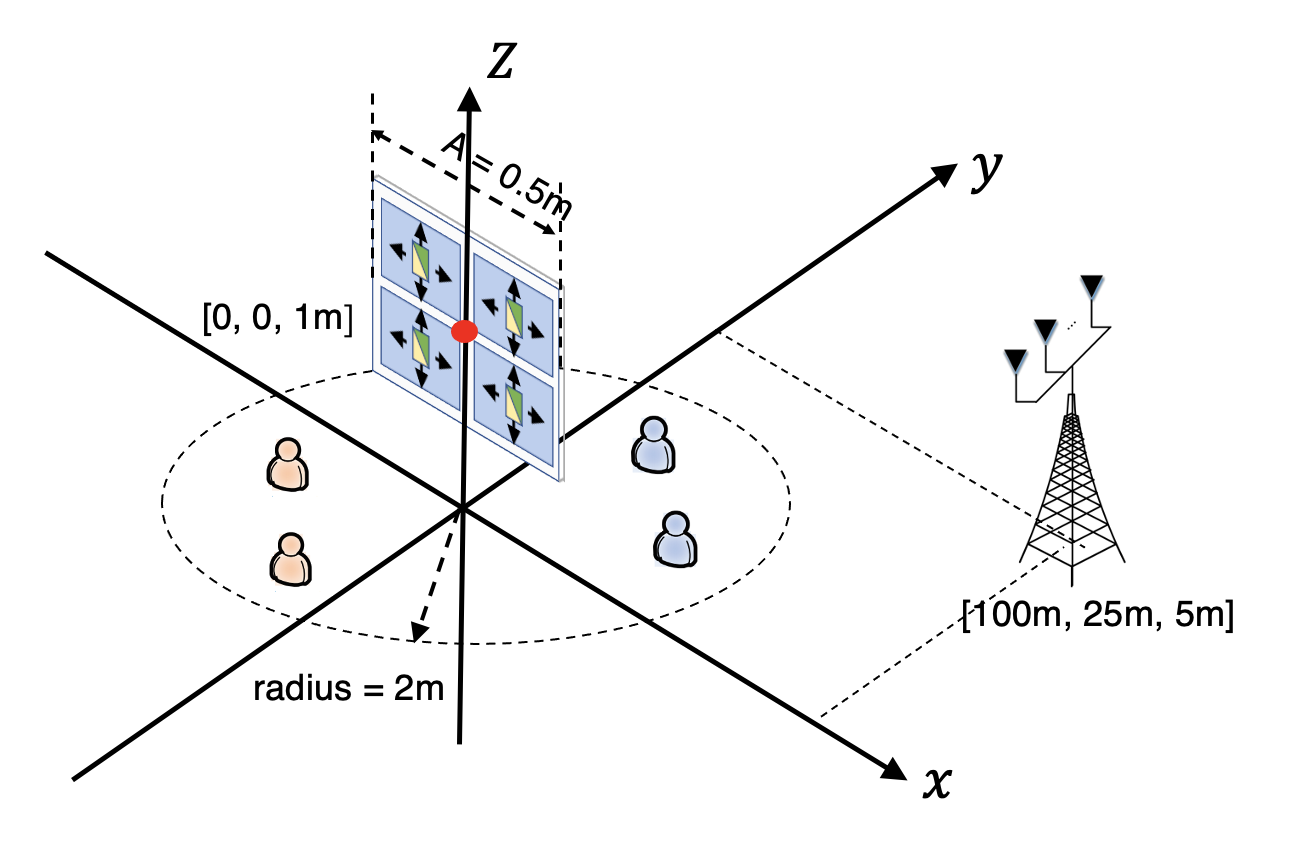}
	\caption{Simulation setup.}
	\label{simulation}
\end{figure}
As shown in Fig. \ref{simulation}, the simulation setup considers a three-dimensional (3-D) coordinate system, where the BS is located at $[100\textup{m}, 25\textup{m}, 5\textup{m}]^T$ and the ME-STARS is deployed perpendicular to the $y$-axis at the origin, with its center coordinates at $[0,0,1\textup{m}]^T$. Additionally, we assume that two users are randomly distributed within the semicircular region of radius 2 meters on each of the T and R regions of the ME-STARS. The main system parameters are set as follows, unless otherwise specified: the number of antennas at the BS is $N=4$, the center communication frequency $f_c = 40$ GHz, and the bandwidth $B = 10$ GHz is distributed over $L=11$ subcarriers. Besides, the ME-STARS is assigned with a square shape of size $A\times A$, and $A=0.5$ m. The allowable maximum transmit power and the noise power are set to be $P_{\max} =15$ W and $\sigma^2=-110$ dBm. The minimum distance to avoid coupling effect is set to be $d_{\min}=\frac{\lambda_c}{2}$. For the proposed algorithm, the convergence threshold and the number of Gaussian randomizations in the inner loop are set to be $\varepsilon_0=10^{-3}$ and $G=200$, respectively. While for the PSO-based method in the outer layer, the number of particles is $I = 20$, and personal and global learning factors are $c_1=c_2=2$. Besides, $\omega_{\max}=1$, $\omega_{\min}=0.2$, and $T=100$. 

Next, we elaborate on the specific element setups of the ME-STARS considered in this paper. 1) \textbf{ME-STARSs with RB mode}: In this scheme, we employ a 2-D movable structure, similar to the movable antenna design in \cite{Zhu_movable}, to deploy the movable elements efficiently. 2) \textbf{ME-STARSs with HB mode:} In this scheme, we directly adopt a structure consisting of $M$ equally spaced horizontal tracks, with each track accommodating a single movable element. \textcolor{black}{Mathematically, we have $Z_{num}=M$, $Z_m=m$, and $\delta_H=\frac{A}{M}$.} This configuration ensures that, as long as $M$ is not excessively large, the horizontal spacing between elements remains greater than the minimum required distance $d_{\min}$. 3) \textbf{ME-STARSs with VB mode:} Similar to the HB mode, we also deploy one movable element per track in this scheme, but the tracks are arranged vertically instead. \textcolor{black}{Here, $X_{num}=M$, $X_m=m$, and $\delta_V=\frac{A}{M}$.} 4) \textbf{ME-STARSs with DB mode:} In this scheme, we introduce diagonal tracks of varying lengths, evenly embedded within the ME-STARS, with one movable element placed on each track. Additionally, to serve as a benchmark for performance comparison, we consider a fixed-element STARS configuration, referred to as \textbf{FP-STARSs}. In this scheme, all elements remain fixed in position, determined by the initial setup of the RB mode. Based on these configurations, the joint beamforming optimization problem at the BS and STARS can be efficiently solved using \textbf{Algorithm \ref{alg:A}}.

\subsection{Convergence of Proposed Algorithm}
\begin{figure}[t]
	\setlength{\abovecaptionskip}{0cm}   
	\setlength{\belowcaptionskip}{0cm}   
	\setlength{\textfloatsep}{7pt}
	\centering
	\includegraphics[width=3.5in]{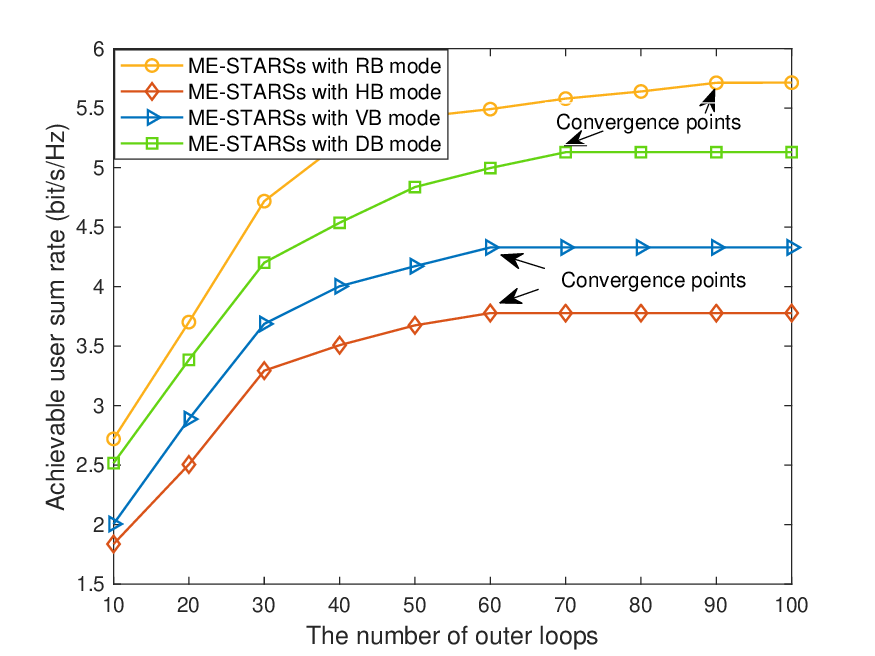}
	\caption{Convergence behavior of \textbf{Algorithm 2}.}
	\label{Convergence}
\end{figure}
Fig. \ref{Convergence} plots the convergence behavior of the proposed algorithm for $M=16$. In particular, we focus on the outer PSO-based \textbf{Algorithm \ref{alg:B}}. As can be observed, the algorithm converges within 100 iterations for all four movement modes, with faster convergence observed for the HB and VB modes and slower convergence for the RB mode. This difference arises because the elements in the HB and VB modes are restricted to 1-D movement, reducing the number of variables and simplifying optimization. In contrast, the RB mode allows 2-D movement, which, while achieving higher performance gains, increases the number of variables and the optimization complexity. Additionally, the hardware implementation of the HB and VB modes avoids coupling effects between different elements, reducing the constraints in the optimization problem. However, the RB mode lacks this advantage, requiring more iterations to meet the coupling constraints. Interestingly, the DB mode achieves both better performance and faster convergence within fewer iterations. This is attributed to its \emph{virtual} 2-D movement, where each element can adjust both its $x$- and $z$-coordinates to optimize performance. However, due to the strong correlation between the $x$- and $z$-coordinates along their diagonal tracks, the optimization variables can effectively be treated as a single parameter, simplifying the optimization process.
\subsection{Evaluation of Normalized Array Gain Versus ME-STARS Movement Modes}
\begin{figure}[]
	\subfigure[$B=4$ GHz.]{\label{array1}
		\includegraphics[width= 3.7in]{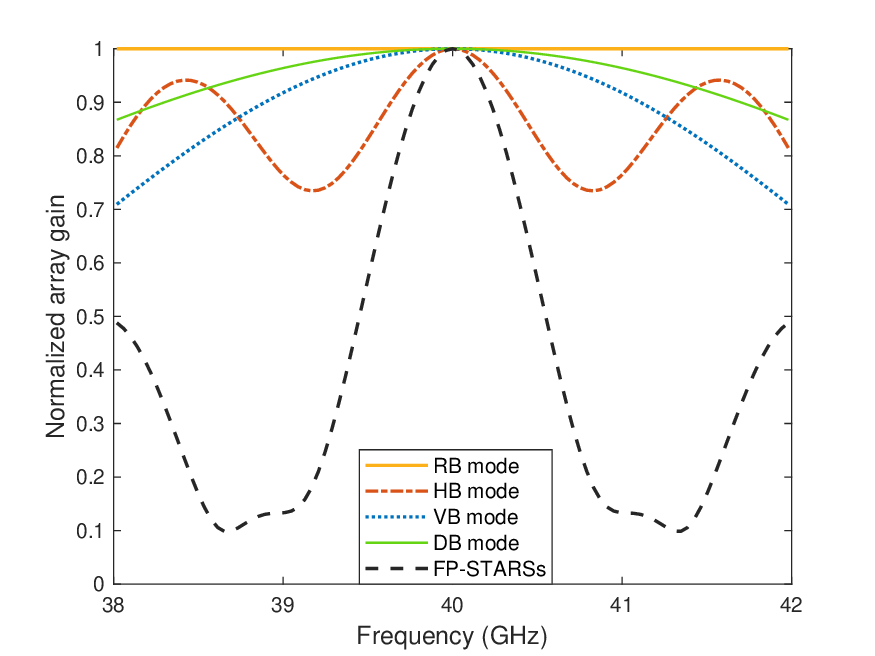}}
	\subfigure[$B=10$ GHz.]{\label{array2}
		\includegraphics[width= 3.7in]{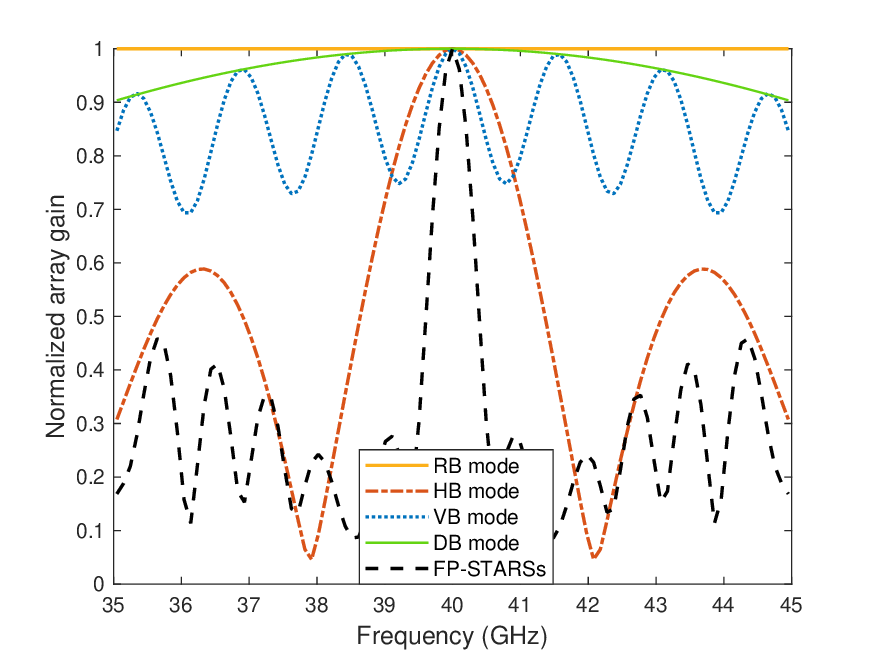}}
	\caption{\textcolor{black}{Evaluation of normalized array gain versus ME-STARS movement modes.} } 
	\label{array}
\end{figure}
Building on the analysis in Fig. \ref{beam_squint}, this subsection continues to target a specific user. Additionally, we adjust the fitness function in \textbf{Algorithm 2} by replacing $\mathcal{R}(\widetilde{\mathbf{S}}_i)$ with the corresponding normalized array gain for this user, computed using equation \eqref{normalized}, to obtain the subsequent results.

Fig. \ref{array} plots the evaluation of normalized array gain versus different ME-STARS movement modes. In particular, we only consider the array gain of the user in region R and with the number of ME-STARS elements $M$ = 8. As observed in Fig. \ref{array1} with $B=4$ GHz, there exists a significant beam squint effect in FP-STARSs. However, this effect can be effectively suppressed by adjusting the position of the elements dynamically, as presented in the figure about the array gain achieved by ME-STARSs. Additionally, the different movable models demonstrated varying levels of suppression. The RB mode achieved complete elimination, while the DB and VB modes provided the maximum possible suppression. In contrast, the HB mode performed the worst. This can be attributed to the fact that the elements in the RB mode have the most flexible mobility and the largest movement range, resulting in the best performance. However, this comes at the cost of increased computational complexity and hardware implementation overhead.

When the bandwidth is extended to 10 GHz, as shown in Fig. \ref{array2}, the performance advantage of the RB mode becomes even more pronounced, maintaining a very high array gain across the entire band. In contrast, both the HB and VB modes exhibit varying degrees of performance degradation. This is because the beam squint effect becomes more pronounced as the bandwidth increases, and the limited 1-D element movement is insufficient to compensate for the performance degradation caused by the wider bandwidth. However, the DB mode demonstrates strong robustness, confirming that it is an excellent potential solution that strikes a balance between hardware implementation and system performance.

\subsection{Achievable Sum User Rate Versus Number of STARS Elements}
\begin{figure}[t]
	\setlength{\abovecaptionskip}{0cm}   
	\setlength{\belowcaptionskip}{0cm}   
	\setlength{\textfloatsep}{7pt}
	\centering
	\includegraphics[width=3.7in]{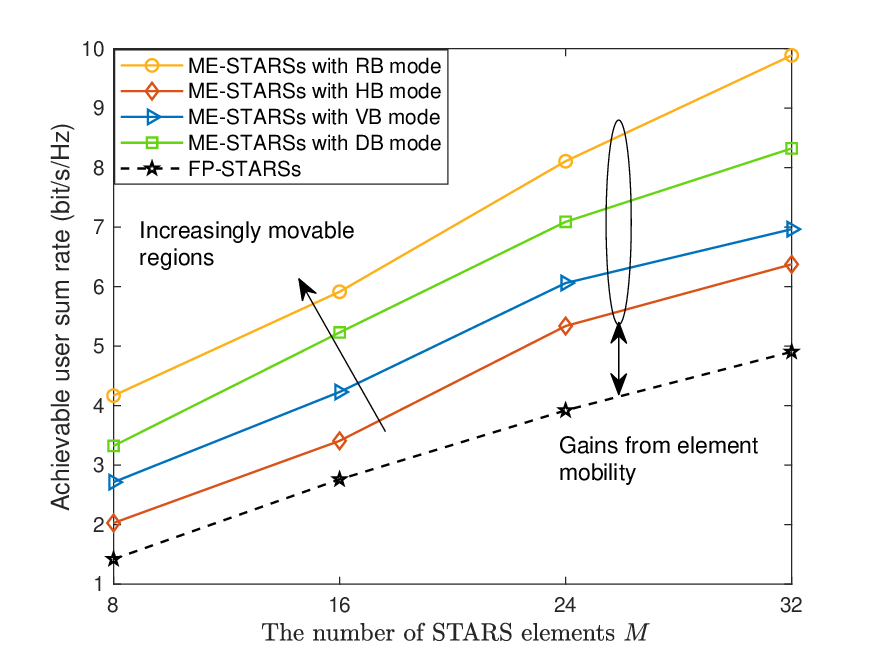}
	\caption{Achievable sum user rate versus the number of STARS elements.}
	\label{elements}
\end{figure}
In Fig. \ref{elements}, we examine the achievable sum user rate versus the number of STARS elements. The results indicate that the performance of all schemes improves with an increasing number of elements $M$, benefiting from the DoFs gained by the additional elements. Moreover, ME-STARSs consistently outperform FP-STARSs due to the advantages of element mobility, with the performance gap widening as $M$ increases. This is expected since a larger $M$ exacerbates the beam squint effect in near-field wideband communications. However, ME-STARSs can effectively mitigate this effect through the flexible movement of their elements, a capability absent in FP-STARSs. This reasoning also explains the performance gaps between various ME-STARS movement modes. As the movable area expands, ME-STARSs are better able to suppress the beam squint effect, resulting in enhanced system performance. Accordingly, modes permitting 2-D movement outperform those restricted to 1-D movement. Among them, the RB mode, which provides the most flexible movement, achieves the best performance. Furthermore, regarding the HB and VB modes, both featuring similar 1-D movable regions, the VB mode outperforms the HB mode. This is because vertically moving elements affect only the elevation-related terms in the array response, whereas horizontally moving elements impact both elevation-related and azimuth-related terms, making it more challenging to find an optimal position to mitigate the beam squint effect.
\subsection{Achievable Sum User Rate Versus Number of Subcarriers}
\begin{figure}[t]
	\setlength{\abovecaptionskip}{0cm}   
	\setlength{\belowcaptionskip}{0cm}   
	\setlength{\textfloatsep}{7pt}
	\centering
	\includegraphics[width=3.7in]{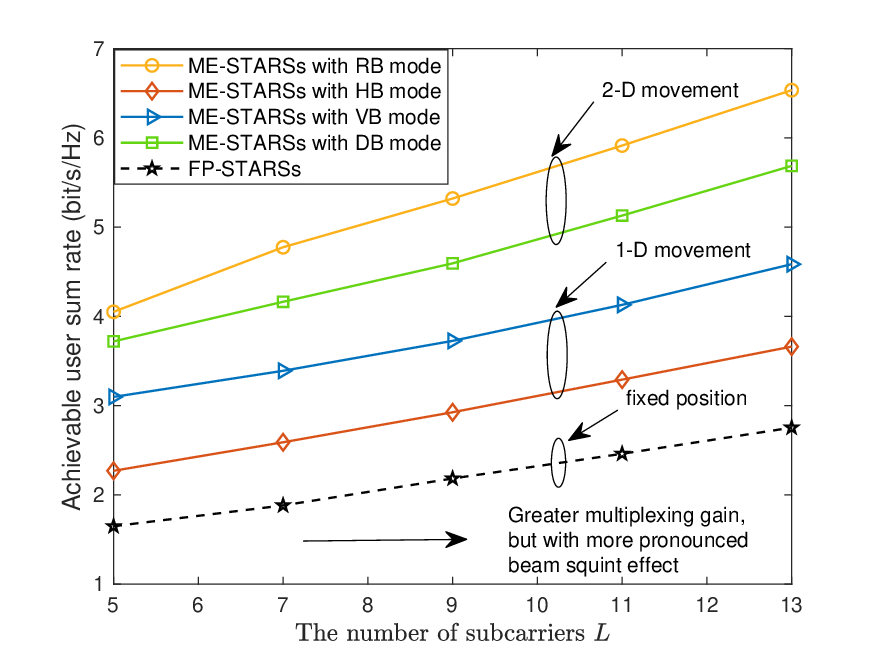}
	\caption{Achievable sum user rate versus the number of subcarriers.}
	\label{subcarriers}
\end{figure}
As shown in Fig. \ref{subcarriers}, we study the achievable sum rate as a function of the number of subcarriers $L$. It can be seen that the achievable performance of all schemes increases as $L$ grows, as more subcarriers provide access to additional spectrum resources and enable better frequency multiplexing. This allows the system to fully leverage multipath effects and frequency-selective fading. However, the performance gap between the schemes widens as $L$. This is because the beam squint effect becomes more pronounced with a larger $L$, necessitating effective mitigation strategies to achieve optimal performance gains. Among all modes, the RB mode excels at addressing this challenge, which explains its growing performance advantage as $L$ increases.

\section{Conclusion}
This study examined the problem of maximizing the sum of achievable user rates in ME-STARS-assisted near-field wideband communications. To address the beam squint effect caused by frequency-independent phase shift designs at the STARS, we proposed a cost-effective ME-STARS architecture and introduced four practical element movement modes. For each mode, we developed a comprehensive framework for the joint optimization of BS precoding, ME-STARS beamforming, and element positioning to achieve the optimization objectives. To tackle the resulting optimization problem, we proposed an efficient two-layer algorithm, leveraging the PSO search method in the outer layer and the BCD optimization framework in the inner layer. Numerical results demonstrated that enabling mobility for the elements in the STARS effectively mitigates the beam squint effect. Specifically, the RB mode achieves perfect mitigation of the beam squint effect, while the DB mode strikes a more balanced trade-off between performance gain and hardware implementation complexity.

\textcolor{black}{Due to the scope of this work, several important challenges in the ME-STARS remain unaddressed. These issues present key opportunities for future research and are discussed below.}
\begin{itemize}
	\item\textcolor{black}{\emph{Practical Implementation Constraints}: While element mobility improves performance, the practical deployment of the ME-STARS requires addressing key challenges such as mechanical accuracy, energy efficiency, and real-time reconfigurability. The design of lightweight and low-power actuation mechanisms for element movement remains an open issue.}
	\item\textcolor{black}{\emph{Optimization of Dynamic Element Movement}: Existing optimization techniques predominantly focus on static or semi-static element positioning in ME-STARS-assisted communications. To facilitate practical applications, it is essential to develop efficient algorithms for real-time dynamic element reconfiguration while considering mobility constraints and system latency.}
	\item\textcolor{black}{\emph{Application for Next Generation Multiple Access (NGMA) in 6G Networks\cite{Mu_function}}: Future research should investigate the potential of movable-STARS in enhancing performance for NGMA in 6G and beyond networks. A deeper theoretical and experimental analysis is needed to understand the interactions between mobile elements, user mobility, and interference management, ensuring efficient and adaptive communication in multi-user and multi-access scenarios. Besides, interplaying ME-STARS and emerging technologies, such as integrated sensing and communications (ISAC), near-field communications (NFC), and simultaneous wireless information and power transfer (SWIPT), is also a direction of interest.}
	\item\textcolor{black}{\emph{AI-Driven Adaptive Control}: Integrating artificial intelligence (AI) and machine learning (ML) techniques into ME-STARS has the potential to enable real-time adaptive control in response to dynamic environmental conditions. Exploring AI-driven optimization frameworks for element positioning and phase shift adaptation presents a promising research direction, offering enhanced flexibility and efficiency in reconfigurable wireless networks.}
\end{itemize}
\balance
\bibliographystyle{IEEEtran} 
\bibliography{movable.bib}

\begin{thebibliography}{10}
\providecommand{\url}[1]{#1}
\csname url@samestyle\endcsname
\providecommand{\newblock}{\relax}
\providecommand{\bibinfo}[2]{#2}
\providecommand{\BIBentrySTDinterwordspacing}{\spaceskip=0pt\relax}
\providecommand{\BIBentryALTinterwordstretchfactor}{4}
\providecommand{\BIBentryALTinterwordspacing}{\spaceskip=\fontdimen2\font plus
\BIBentryALTinterwordstretchfactor\fontdimen3\font minus
  \fontdimen4\font\relax}
\providecommand{\BIBforeignlanguage}[2]{{%
\expandafter\ifx\csname l@#1\endcsname\relax
\typeout{** WARNING: IEEEtran.bst: No hyphenation pattern has been}%
\typeout{** loaded for the language `#1'. Using the pattern for}%
\typeout{** the default language instead.}%
\else
\language=\csname l@#1\endcsname
\fi
#2}}
\providecommand{\BIBdecl}{\relax}
\BIBdecl

\bibitem{Zhu_conference}
G.~Zhu, X.~Mu, L.~Guo, H.~Ao, and S.~Xu, ``Beam squint mitigation in
  movable-element {STARS}-aided near-field communications,'' in \emph{Proc.
  {IEEE/CIC} Int. Conf. Commun. China ({ICCC})}, Aug. 2025, submitted.

\bibitem{6G_high-frequency}
W.~Saad, M.~Bennis, and M.~Chen, ``A vision of {6G} wireless systems:
  Applications, trends, technologies, and open research problems,''
  \emph{{IEEE} Netw.}, vol.~34, no.~3, pp. 134--142, May/Jun. 2020.

\bibitem{6G_application}
H.~Tataria, M.~Shafi, A.~F. Molisch, M.~Dohler, H.~Sjöland, and F.~Tufvesson,
  ``{6G} wireless systems: Vision, requirements, challenges, insights, and
  opportunities,'' \emph{Proc. IEEE}, vol. 109, no.~7, pp. 1166--1199, Jul.
  2021.

\bibitem{RIS_survey}
Y.~Liu, X.~Liu, X.~Mu, T.~Hou, J.~Xu, M.~Di~Renzo, and N.~Al-Dhahir,
  ``Reconfigurable intelligent surfaces: Principles and opportunities,''
  \emph{{IEEE} Commun. Surv. Tut.}, vol.~23, no.~3, pp. 1546--1577, 3rd Quart.
  2021.

\bibitem{Huang_EE}
C.~Huang, A.~Zappone, G.~C. Alexandropoulos, M.~Debbah, and C.~Yuen,
  ``Reconfigurable intelligent surfaces for energy efficiency in wireless
  communication,'' \emph{{IEEE} Trans. Wireless Commun.}, vol.~18, no.~8, pp.
  4157--4170, Aug. 2019.

\bibitem{Mu_star}
X.~Mu, Y.~Liu, L.~Guo, J.~Lin, and R.~Schober, ``Simultaneously transmitting
  and reflecting ({STAR}) {RIS} aided wireless communications,'' \emph{{IEEE}
  Trans. Wireless Commun.}, vol.~21, no.~5, pp. 3083--3098, May 2022.

\bibitem{Mu_survey}
X.~Mu, J.~Xu, Z.~Wang, and N.~Al-Dhahir, ``Simultaneously transmitting and
  reflecting surfaces for ubiquitous next generation multiple access in {6G}
  and beyond,'' \emph{Proc. IEEE}, vol. 112, no.~9, pp. 1346--1371, Sep. 2024.

\bibitem{Liu_NF_survey}
Y.~Liu, C.~Ouyang, Z.~Wang, J.~Xu, X.~Mu, and A.~L. Swindlehurst, ``Near-field
  communications: A comprehensive survey,'' \emph{{IEEE} Commun. Surv. Tut.},
  pp. 1--1, early access, 2024.

\bibitem{Far_field_beamforming}
R.~W. Heath, N.~González-Prelcic, S.~Rangan, W.~Roh, and A.~M. Sayeed, ``An
  overview of signal processing techniques for millimeter wave {MIMO}
  systems,'' \emph{{IEEE} J. Sel. Topics Signal Process.}, vol.~10, no.~3, pp.
  436--453, Apr. 2016.

\bibitem{Near_field_beamforming}
H.~Zhang, N.~Shlezinger, F.~Guidi, D.~Dardari, M.~F. Imani, and Y.~C. Eldar,
  ``Beam focusing for near-field multiuser {MIMO} communications,''
  \emph{{IEEE} Trans. Wireless Commun.}, vol.~21, no.~9, pp. 7476--7490, Sep.
  2022.

\bibitem{Zhang_NF}
H.~Zhang, N.~Shlezinger, F.~Guidi, D.~Dardari, and Y.~C. Eldar, ``{6G} wireless
  communications: From far-field beam steering to near-field beam focusing,''
  \emph{{IEEE} Commun. Mag.}, vol.~61, no.~4, pp. 72--77, Apr. 2023.

\bibitem{Xu_STAR-RIS_NF}
J.~Xu, X.~Mu, and Y.~Liu, ``Exploiting {STAR-RISs} in near-field
  communications,'' \emph{{IEEE} Trans. Wireless Commun.}, vol.~23, no.~3, pp.
  2181--2196, Mar. 2024.

\bibitem{Mu_RIS_NF}
X.~Mu, J.~Xu, Y.~Liu, and L.~Hanzo, ``Reconfigurable intelligent surface-aided
  near-field communications for {6G}: Opportunities and challenges,''
  \emph{{IEEE} Veh. Technol. Mag.}, vol.~19, no.~1, pp. 65--74, Mar. 2024.

\bibitem{wideband_BS}
B.~Wang, F.~Gao, S.~Jin, H.~Lin, and G.~Y. Li, ``Spatial- and
  frequency-wideband effects in millimeter-wave massive {MIMO} systems,''
  \emph{{IEEE} Trans. Signal Process.}, vol.~66, no.~13, pp. 3393--3406, Jul.
  2018.

\bibitem{Zhang_NF_wideband}
Y.~Zhang, H.~Zhang, S.~Xiao, W.~Tang, and Y.~C. Eldar, ``Near-field wideband
  secure communications: An analog beamfocusing approach,'' \emph{{IEEE} Trans.
  Signal Process.}, vol.~72, pp. 2173--2187, 2024.

\bibitem{Wang_wideband}
Z.~Wang, X.~Mu, and Y.~Liu, ``Beamfocusing optimization for near-field wideband
  multi-user communications,'' \emph{{IEEE} Trans. Commun.}, vol.~73, no.~1,
  pp. 555--572, Jan. 2025.

\bibitem{Guo_NF_wideband}
Y.~Guo, Y.~Zhang, Z.~Wang, and Y.~Liu, ``Wideband beamforming for near-field
  communications with circular arrays,'' \emph{{IEEE} Trans. Wireless Commun.},
  vol.~23, no.~12, pp. 19\,065--19\,082, Dec. 2024.

\bibitem{Cui_NF_wideband}
M.~Cui and L.~Dai, ``Near-field wideband beamforming for extremely large
  antenna arrays,'' \emph{{IEEE} Trans. Wireless Commun.}, vol.~23, no.~10, pp.
  13\,110--13\,124, Oct. 2024.

\bibitem{Cui_NF_wideband_traning}
M.~Cui, L.~Dai, Z.~Wang, S.~Zhou, and N.~Ge, ``Near-field rainbow: Wideband
  beam training for {XL-MIMO},'' \emph{{IEEE} Trans. Wireless Commun.},
  vol.~22, no.~6, pp. 3899--3912, Jun. 2023.

\bibitem{Myers_NF_wideband}
N.~J. Myers and R.~W. Heath, ``{InFocus}: A spatial coding technique to
  mitigate misfocus in near-field {LoS} beamforming,'' \emph{{IEEE} Trans.
  Wireless Commun.}, vol.~21, no.~4, pp. 2193--2209, Apr. 2022.

\bibitem{Cheng_RIS_NF_wideband}
Y.~Cheng, C.~Huang, W.~Peng, M.~Debbah, L.~Hanzo, and C.~Yuen, ``Achievable
  rate optimization of the {RIS}-aided near-field wideband uplink,''
  \emph{{IEEE} Trans. Wireless Commun.}, vol.~23, no.~3, pp. 2296--2311, Mar.
  2024.

\bibitem{Wang_RIS_NF_wideband}
J.~Wang, J.~Xiao, Y.~Zou, W.~Xie, and Y.~Liu, ``Wideband beamforming for {RIS}
  assisted near-field communications,'' \emph{{IEEE} Trans. Wireless Commun.},
  vol.~23, no.~11, pp. 16\,836--16\,851, Nov. 2024.

\bibitem{Yang_RIS_NF_wideband}
S.~Yang, C.~Xie, W.~Lyu, B.~Ning, Z.~Zhang, and C.~Yuen, ``Near-field channel
  estimation for extremely large-scale reconfigurable intelligent surface
  ({XL-RIS})-aided wideband mmwave systems,'' \emph{{IEEE} J. Sel. Areas
  Commun.}, vol.~42, no.~6, pp. 1567--1582, Jun. 2024.

\bibitem{Hao_beam_squint}
W.~Hao, X.~You, F.~Zhou, Z.~Chu, G.~Sun, and P.~Xiao, ``The far-/near-field
  beam squint and solutions for {THz} intelligent reflecting surface
  communications,'' \emph{{IEEE} Trans. Veh. Technol.}, vol.~72, no.~8, pp.
  10\,107--10\,118, Aug. 2023.

\bibitem{Li_RIS_NF_wideband}
Z.~Li, Z.~Gao, and T.~Li, ``Sensing user's channel and location with terahertz
  extra-large reconfigurable intelligent surface under hybrid-field beam squint
  effect,'' \emph{{IEEE} J. Sel. Topics Signal Process.}, vol.~17, no.~4, pp.
  893--911, Jul. 2023.

\bibitem{Li_RIS_wideband}
J.~Li, S.~Zhang, Z.~Li, J.~Ma, and O.~A. Dobre, ``User sensing in {RIS}-aided
  wideband mmwave system with beam-squint and beam-split,'' \emph{{IEEE} Trans.
  Commun.}, pp. 1--1, early access. 2024.

\bibitem{Wong_Fluid}
K.-K. Wong, A.~Shojaeifard, K.-F. Tong, and Y.~Zhang, ``Fluid antenna
  systems,'' \emph{{IEEE} Trans. Wireless Commun.}, vol.~20, no.~3, pp.
  1950--1962, Mar. 2021.

\bibitem{Wong_Fluid_access}
K.-K. Wong and K.-F. Tong, ``Fluid antenna multiple access,'' \emph{{IEEE}
  Trans. Wireless Commun.}, vol.~21, no.~7, pp. 4801--4815, Jul. 2022.

\bibitem{Zhu_movable}
L.~Zhu, W.~Ma, and R.~Zhang, ``Modeling and performance analysis for movable
  antenna enabled wireless communications,'' \emph{{IEEE} Trans. Wireless
  Commun.}, vol.~23, no.~6, pp. 6234--6250, Jun. 2024.

\bibitem{Ma_movable}
W.~Ma, L.~Zhu, and R.~Zhang, ``{MIMO} capacity characterization for movable
  antenna systems,'' \emph{{IEEE} Trans. Wireless Commun.}, vol.~23, no.~4, pp.
  3392--3407, Apr. 2024.

\bibitem{Yan_STAR_wideband}
W.~Yan, W.~Hao, G.~Sun, C.~Huang, and Q.~Wu, ``Wideband beamforming for
  {STAR-RIS}-assisted {THz} communications with three-side beam split,''
  \emph{{IEEE} Trans. Commun.}, pp. 1--1, early access, 2024.

\bibitem{Rayleigh_distance}
C.~A. Balanis, \emph{Antenna theory: analysis and design}.\hskip 1em plus 0.5em
  minus 0.4em\relax Hoboken, NJ, USA: Wiley, 2016.

\bibitem{Wu_estimation}
C.~Wu, C.~You, Y.~Liu, X.~Gu, and Y.~Cai, ``Channel estimation for
  {STAR-RIS}-aided wireless communication,'' \emph{{IEEE} Wireless Commun.
  Lett.}, vol.~26, no.~3, pp. 652--656, Mar. 2022.

\bibitem{Xiao_estimation}
Z.~Xiao, S.~Cao, L.~Zhu, Y.~Liu, B.~Ning, X.-G. Xia, and R.~Zhang, ``Channel
  estimation for movable antenna communication systems: A framework based on
  compressed sensing,'' \emph{{IEEE} Trans. Wireless Commun.}, vol.~23, no.~9,
  pp. 11\,814--11\,830, Sep. 2024.

\bibitem{wideband_power}
J.~P. González-Coma, J.~Rodríguez-Fernández, N.~González-Prelcic,
  L.~Castedo, and R.~W. Heath, ``Channel estimation and hybrid precoding for
  frequency selective multiuser mmwave {MIMO} systems,'' \emph{{IEEE} J. Sel.
  Topics Signal Process.}, vol.~12, no.~2, pp. 353--367, May 2018.

\bibitem{Zhu_SWIPT}
G.~Zhu, X.~Mu, L.~Guo, A.~Huang, and S.~Xu, ``Robust resource allocation for
  {STAR-RIS} assisted {SWIPT} systems,'' \emph{{IEEE} Trans. Wireless Commun.},
  vol.~23, no.~6, pp. 5616--5631, Jun. 2024.

\bibitem{Huang_SDP}
Y.~Huang and D.~P. Palomar, ``Rank-constrained separable semidefinite
  programming with applications to optimal beamforming,'' \emph{IEEE
  Transactions on Signal Processing}, vol.~58, no.~2, pp. 664--678, Feb. 2010.

\bibitem{Zhu_FP}
G.~Zhu, X.~Mu, L.~Guo, A.~Huang, and S.~Xu, ``Enhancing user fairness in
  wireless powered communication networks with {STAR-RIS},'' \emph{IEEE
  Internet of Things J.}, vol.~12, no.~3, pp. 2659--2673, Feb. 2025.

\bibitem{cvx}
M.~Grant and S.~Boyd, ``{CVX}: {MATLAB} software for disciplined convex
  programming, version 2.1,'' [Online]. Available:\url{http://cvxr.com/cvx},
  2014.

\bibitem{Rank-one}
X.~Mu, Y.~Liu, L.~Guo, J.~Lin, and N.~Al-Dhahir, ``Exploiting intelligent
  reflecting surfaces in {NOMA} networks: Joint beamforming optimization,''
  \emph{{IEEE} Trans. Wireless Commun.}, vol.~19, no.~10, pp. 6884--6898, Oct.
  2020.

\bibitem{Wu_Gaussian}
Q.~Wu and R.~Zhang, ``Intelligent reflecting surface enhanced wireless network
  via joint active and passive beamforming,'' \emph{{IEEE} Trans. Commun.},
  vol.~18, no.~11, pp. 5394--5409, Nov. 2019.

\bibitem{PSO}
Y.~Shi and R.~Eberhart, ``A modified particle swarm optimizer,'' in \emph{in
  Proc. IEEE Int. Conf. Evol. Comput. World Congr. Comput. Intel,}, Anchorage,
  AK, US 1998, pp. 69--73.

\bibitem{Xiao_PSO}
Z.~Xiao, X.~Pi, L.~Zhu, X.-G. Xia, and R.~Zhang, ``Multiuser communications
  with movable-antenna base station: Joint antenna positioning, receive
  combining, and power control,'' \emph{{IEEE} Trans. Wireless Commun.},
  vol.~23, no.~12, pp. 19\,744--19\,759, Dec. 2024.

\bibitem{SDR}
Z.-Q. Luo, W.-K. Ma, A.~M.-C. So, Y.~Ye, and S.~Zhang, ``Semidefinite
  relaxation of quadratic optimization problems,'' \emph{IEEE Signal Processing
  Mag.}, vol.~27, no.~3, pp. 20--34, May 2010.

\bibitem{PSO_convergence}
W.~Qian and M.~Li, ``Convergence analysis of standard particle swarm
  optimization algorithm and its improvement,'' \emph{Soft Computing}, vol.~22,
  pp. 4047--4070, 2018.

\bibitem{Mu_function}
X.~Mu, Z.~Wang, and Y.~Liu, ``Simultaneously transmitting and reflecting
  surfaces ({STARS}) for multi-functional {6G},'' \emph{{IEEE} Netw.}, vol.~39,
  no.~1, pp. 47--55, Jan. 2025.

\end{thebibliography}
\end{document}